\documentclass[12pt]{article}

\usepackage{amscd}
\usepackage{amsmath}

\begin{document}

\newcommand{\N}{N\raise.7ex\hbox{\underline{$\circ $}}$\;$}

\thispagestyle{empty}

\begin{center}

{\bf On 15-component  theory of a charged spin-1  particle\\ with
 polarizability in  Coulomb and Dirac monopole fields
}
\end{center}

\begin{center}

{\bf   Red'kov V.M.\footnote{E-mail: redkov@dragon.bas-net.by},
Tokarevskaya N.G., Kisel V.V.  }
\\ Belarus  National  Academy of Sciences,
B.I. Stepanov's Institute of Physics \\
Belarus State Maksim Tank's Pedagogical University

\end{center}

\begin{quotation}

The problem  of a spin 1 charged particle with electromagnetic
polarizability,
obeying a generalized 15-component quantum mechanical equation, is  investigated in
presence of the external  Coulomb potential.
With the use of the Wigner's functions techniques, separation of
variables in the spherical  tetrad basis is done and the 15-component  radial system is given.
It is shown that there exists a class of quantum states for which
the additional characteristics,  polarizability, does not
manifest itself anyhow; at this  the  energy spectrum of the
system coincides with the known spectrum of the  scalar particle. For $j=0$ states,  a
2-order   differential equation is derived, it contains an additional potential term
$r^{-4}$.

In analogous approach wave functions of the generalized particle are examined in presence of external Dirac monopole
field. It is shown that  there exists one special state with minimal conserved quantum number $j_{min}$.
It this solution, first,  the polarizability does not exhibits itself, and second,  this solution provides us with
analogue of the known peculiar solution in the theory of the Dirac particle in external monopole field,
 which can describe certain bound quantum mechanical state.

Analysis of the usual vector particle in external Coulomb  potential, additional to that existing in the
literature, is given. It is shown that at $j=0$ some bound states will arise. The corresponding energy spectrum is found.

\end{quotation}

\section{Basic equation and notation}

Initial equation for a 15-component wave function of the particle with electric charge
and polarizability [1-8]  has in matrix formalism the form
\begin{eqnarray}
(\Gamma^{a} \partial_{a}   -  m ) \Psi = 0 \;, \;\; \Psi = \left |
\begin{array}{l} C \\ C_{l}  \\    \Phi_{l} \\ \Phi_{mn}
\end{array} \right |                \; , \;\;
\Gamma^{a} =
\left | \begin{array}{llll}
0  &  G^{a} &  0  &  0  \\
0  &  0  &  0  &  K^{a} \\
\sigma \Delta^{a}  &  0  &  0 &  K^{a} \\
0 &  0  &  \Lambda ^{a}  &  0
\end{array} \right | \; ,
\label{1.1}
\end{eqnarray}

\noindent  here blocks of dimensions
$1\times4$, $4\times1$, $4\times6$ и $6\times4$ respectively  are used:
$$
(G^{a})_{(0)}^{\;\;\;\;\;k}=g^{ak} \; ,
\qquad
(\Delta^{a})_{\;\;\;\;n}^{(0)}=\delta_{n}^{a} \; ,
$$$$
(K_{a})_{n}^{\;\;\;kl} =
- g^{ak} \; \delta_{n}^{l} + g^{al} \; \delta_{n}^{k}, \;\;
(\Lambda^{a})_{\;\;\;nb}^{k}=  \delta^{a}_{n} \; \delta^{k}_{b}  -
                   \delta^{k}_{n}  \; \delta^{a}_{b}  \; .
$$

\noindent The $\sigma$ stands for a free parameter of the model. It concerns with an
additional characteristic of $S=1$ particle manifesting itself in external electromagnetic and
gravitational  fields.

As was shown in [1],  equation (1.1) can be extended to general
relativity case  [9-12] in the following way
\begin{eqnarray}
[ \; \Gamma ^{\alpha }(x)\; ( \partial_{\alpha} \;  +  \;
B_{\alpha }(x) ) \; - m \;  ] \;\Psi  (x)  = 0 \; ,
\label{1.3}
\end{eqnarray}

\noindent where
\begin{eqnarray}
\Gamma ^{\alpha }(x) = \Gamma ^{a} e ^{\alpha }_{(a)}(x) \; , \;
B_{\alpha }(x) =
{1 \over 2}\; J^{ab} e ^{\beta }_{(a)}\nabla _{\alpha }( e_{(b)\beta }) \; .
\label{1.4}
\end{eqnarray}

\noindent
In the present work, some technical possibilities in equations
(\ref{1.3}) underlying   by the used tetrad formalism will be  widely
exploited in studying behavior of the generalized vector particle
in presence of Coulomb and Dirac monopole fields. In that sense,
present Section helps us prepare for detailed treatment of these
two problems.

Let us consider equation (\ref{1.3}) in Minkowski space assuming the use of spherical coordinates
and a diagonal tetrad:
\begin{eqnarray}
dS^{2}= c^{2} dt^{2}-dr^{2}-r^{2}(d\theta^{2}+\sin^{2}{\theta}d\phi^{2})=
g_{\alpha\beta}d x^{\alpha} d x^{\beta} \; ,
\nonumber
\\
x^{\alpha}=(c t, r, \theta, \phi) \; , \;
g_{\alpha\beta}=\left | \begin{array}{cccc}
                  1  &  0       &  0       &   0 \\
                  0  &  -1      &  0       &   0 \\
                  0  &  0       &  -r^{2}  &  0  \\
                  0  &  0       &  0   &   -r^{2}\sin^{2}{\theta}
                        \end {array}
                \right | ,
\nonumber
\\
e^{\alpha}_{(0)}=(1, 0, 0, 0), \qquad e^{\alpha}_{(3)}=(0, 1, 0, 0),
\nonumber
\\
e^{\alpha}_{(1)}=(0, 0, \frac {1}{r}, 0), \qquad
e^{\alpha}_{(2)}=(1, 0, 0, \frac{1}{r \sin \theta})  \; .
\label{1.5}
\end{eqnarray}

\noindent
In tetrad  (\ref{1.5}), equation (1.3) will looks
\begin{eqnarray}
\left [\; \Gamma^{0} \partial_{0} \; + \;  \Gamma^{3}\partial_{r} +
\frac{\Gamma^{1} J^{31} + \Gamma^{2} J^{32} } {r}  \; + \;
\frac{1}{r}\; \Sigma_{\theta,\phi } \; - \; m \; \right ] \; \Psi (x) = 0 \; ,
\label{1.10a}
\end{eqnarray}

\noindent
where $\Sigma_{\theta,\phi}$ designates   $\theta, \phi$ - dependent operator
\begin{eqnarray}
\Sigma_{\theta,\phi} = \Gamma^{1} \; \partial_{\theta} \; + \; \Gamma^{2} \;
\frac{\partial_{\phi} + \cos{\theta}J^{12}}{\sin{\theta}} \; .
\label{1.10b}
\end{eqnarray}

We will  need explicit representation for  $\Gamma^{a}$-matrices.
With the use of notation
\begin{eqnarray}
\vec{e}_{1}=\left | \begin{array}{rrr}
      1  &  0  &     0
         \end {array} \right | , \qquad
\vec{e} _{2}=\left | \begin{array}{rrr}
      0  &  1  &     0
         \end {array} \right | , \qquad
\vec{e}_{3} =\left | \begin{array}{rrr}
      0  &  0  &     1
         \end {array} \right | \; ,
\nonumber
\\
\vec{e}^{\;\;t}_{1}=\left | \begin{array}{c}
         1  \\  0  \\     0
         \end {array}  \right | ,   \qquad
\vec{e}^{\;\;t}_{2} = \left | \begin{array}{c}
         0  \\  1  \\     0
         \end {array} \right |, \qquad
\vec{e}^{\;\;t}_{3} =\left | \begin{array}{c}
         0  \\  0  \\ 1
         \end {array} \right | \; ,
\nonumber
\\
\tau_{1}=
\left | \begin{array}{rrr}
      0  &  0  &     0\\
      0  &  0  &    -1\\
      0  &  1  &     0
         \end {array} \right | , \;
\tau_{2}=
\left | \begin{array}{rrr}
      0  &  0  &     1\\
      0  &  0  &     0\\
     -1  &  0  &     0
         \end {array} \right |,\;
\tau_{3}=
\left | \begin{array}{rrr}
      0  &  -1 &     0\\
      1  &  0  &     0\\
      0  &  0  &     0
         \end {array} \right | \;,
\nonumber
\end{eqnarray}

\noindent
\noindent
 the $\Gamma^{a}$ look as
\begin{eqnarray}
\Gamma^{0} =
\left | \begin{array}{ccccccc}
      0  &  1  &    \vec{0}  &  0       &  \vec{0}       &   \vec{0}  &  \vec{0} \\
      0  &  0  &    \vec{0}  &  0       &  \vec{0 }      &   \vec{0}   & \vec{0} \\
      \vec{0}^{\;t}   &  \vec{0}^{\;t}  &    0 &   \vec{0}^{\;t}     &  0       &  -I  & 0\\
\sigma   &  0  &    \vec{0}  &  0       &  \vec{0}        &   \vec{0}  & \vec{0} \\
      \vec{0}^{\;t}  &  \vec{0}^{\;t}   &    0 &  \vec{0}^{\;t}  &  0       &  -I  & 0\\
      \vec{0}^{\;t}    &  \vec{0}^{\;t} &    0 &  \vec{0}^{\;t}  &  I       &   0  & 0\\
      \vec{0}^{\;t}  &   \vec{0}^{\;t}  &    0 &  \vec{0}^{\;t}  &  0       &   0  & 0
\end {array}
\right |,
\nonumber
\end{eqnarray}
\begin{eqnarray}
\Gamma^{i} =
\left | \begin{array}{ccccccc}
      0  &  0  &   -\vec{e}_{i} &  0       &  \vec{0}       &  \vec{0}         & \vec{0}      \\
      0  &  0  &    \vec{0}     &  0       &  \vec{0}       &   - \vec{e}_{i}  & \vec{0}       \\
      \vec{0}^{\;t}  &  \vec{0}^{\;t}  &    0     &   \vec{0}^{\; t}        &  0       &   0    & -\tau_{i}\\
      0  &  0  &    \vec{0}    &  0       &  \vec{0}       &   - \vec{e}_{i}  & \vec{0}       \\
      \sigma  \vec{e}^{\;\;t}_{i}  &  \vec{0}^{\;t}  &    0  &   \vec{0}^{\;t}   &  0       &   0      & -\tau_{i} \\
      \vec{0}^{\;t}   &  \vec{0}^{\;t}  &    0     &  - \vec{e}^{\;t}_{i}  &  0       &   0      & 0       \\
      \vec{0}^{\;t}  &  \vec{0}^{\;t}  &    0     &  \vec{0}^{\;t}       &  \tau_{i}&   0      & 0
\end {array} \right | .
\nonumber
\end{eqnarray}

Also we will be needing generators of the Lorentz group representations involved:
$$
J^{ab} =
\left |
 \begin{array}{cccc}
0  &  0      &  0       &  0  \\
0  &  V^{ab} &  0       &  0  \\
0  &  0      &  V^{ab}  &  0  \\
0  &  0      &  0       & (V\otimes V)^{ab}
\end{array} \right |    \; ,
$$

\noindent Vector generators are
$$
(V^{ab})_{k}^{\;\; l}=
-g^{al}\delta^{b}_{k}+g^{bl}\delta^{a}_{k} \;, \qquad
(V^{23})_{k}^{\;\; l}=
\left | \begin{array}{rrrr}
0  &  0      &  0       &  0  \\
0  &  0      &  0       &  0  \\
0  &  0      &  0       &  -1  \\
0  &  0      &  1       &  0
\end{array}
 \right | =
\left | \begin{array}{cc}
0  &    0  \\
0  &    \tau_{1}
\end{array}
 \right | \; ,
$$

\noindent and so on. Tensor generators are  (index combinations $01 , \;   02, \; 03 , \; 23 , \; 31 , \; 12 $
are used)
$$
[(V \otimes V)^{ab}]_{mn}^{\;\;\;\;sp}=
\; (-g^{as} \; \delta^{b}_{m} \; + \; g^{bs} \; \delta_{m}^{a}) \; \delta_{n}^{p} \; + \;
\delta^{s}_{m} \; (-g^{ap} \; \delta^{b}_{n} \; + \; g^{bp} \; \delta_{n}^{a}) \;  \;
$$
$$
[(V \otimes V)^{23}]_{mn}^{\;\;\;\;sp}=
\left | \begin{array}{rrrrrrr}
      0  & 0  &  0  & 0  &  0       & 0\\
      0  & 0  &  -1 & 0  &  0       & 0\\
      0  & 1  &  0  & 0  &  0       & 0\\
      0  & 0  &  0  & 0  &  0       & 0\\
      0  & 0  &  0  & 0  &  0       & -1\\
      0  & 0  &  0  & 0  &  1       & 0\\
\end {array}
\right |=
\left | \begin{array}{cc}
      \tau_{1} & 0        \\
         0     & \tau_{1}
\end {array} \right | \; ,
$$

\noindent and so on.
Total momentum operators in Cartesian basis has the conventional form
\begin{eqnarray}
J_{k} = l_{k} + S_{k} \; ,
\;\;
S_{1} = i J^{23} \; , \;
S_{2} = i J^{31} , \; S_{3} = i J^{12} \; ,
\nonumber
\\
S_{k} =  i \;[ 0 \oplus ( 0 \oplus \tau_{k} ) \oplus ( 0 \oplus \tau_{k} )
\oplus ( \tau_{k} \oplus \tau_{k}) \; ]
\nonumber
\end{eqnarray}

One should have their representation in spherical tetrad basis. To this end, taking the tetrad
transformation law
$$
e^{'\alpha}_{(a)} = {\partial x^{'\alpha} \over \partial x^{\beta}} \;
L_{a}^{\;\;b}  \; e_{(b)}^{\;\;\;\beta} \; , \;
$$

\noindent where Lorentz transformation is reduced to  a pure rotation
$$
O_{i}^{\;\;j}(\theta, \phi)
= \left |  \begin{array}{ccc}
\cos \theta \cos \phi  &  \cos \theta \sin \phi  & - \sin \theta \\
- \sin \phi  &  \cos \phi    &  0  \\
\sin \theta \cos \phi  &  \sin \theta \sin \phi & \cos \theta
\end{array} \right |\; .
$$

\noindent
we arrive at the way to connect Cartesian wave function  $\Psi$  with spherical wave
function  $\Psi '$:
$$
\Psi '(x)=S \; \Psi (x) \; , \;\;
S (\theta, \phi)  = 1 \oplus ( 1 \oplus O) ( 1 \oplus O)
 \oplus ( O \oplus O)\; ]
 $$

\noindent
With the help of the same$S$ one transforms operators
$J_{a}=l_{a}+S_{a},\;\; a=1,\;2,\;3\;$:
$$
J'_{a}=SJ_{a}S^{-1}.
\eqno(1.20)
$$

\noindent Taking in mind the known formulas
\begin{eqnarray}
\l_{1}=i\;(\sin \phi \;\partial_{\theta}\;+\;ctg \;\theta \;\cos \phi \;\partial_{\phi}),
\nonumber
\\
\l_{2}=i\;(-\cos \phi \;\partial_{\theta}\;+\;ctg \;\theta \;\sin \phi \;
\partial_{\phi}),
\qquad
\l_{3}=-i\;\partial_{\phi} \;
\nonumber
\end{eqnarray}

\noindent
and two intermediate ones
\begin{eqnarray}
O\;i\;\partial_{\theta}\;O^{-1}\;=\;i\;\tau_{2}\;,
\qquad
O\;i\;\partial_{\phi}\;O^{-1}\;=
\;i\;(\cos \theta \;\tau_{3}\; -\; \sin \theta \;\tau_{1})\; ,
\nonumber
\end{eqnarray}

\noindent for  $l'_{a}=O\;l_{a}\;O^{-1}$ one derives
\begin{eqnarray}
l'_{1}=l_{1}-i\;(\cos \theta \;\cos \phi \;\tau_{1}+
\;\sin \phi \;\tau_{2}\;+\;\frac{\cos^{2}\theta}{\sin \theta}\;\cos \phi
\;\tau_{3})\;,
\nonumber
\\
l'_{2}=l_{2}\;-\;i\;(\cos \theta \;\sin \phi \;\tau_{1}-
\;\cos \phi \;\tau_{2}\;+\;\frac{\cos^{2}\theta}{\sin \theta}\;\sin \phi \;
\tau_{3})\;,
\nonumber
\\
l'_{3}=l_{3}+\;i\;(\sin \phi \;\tau_{1}-
\;\cos \theta \;\tau_{3})\; ,
\nonumber
\end{eqnarray}

and additionally  for $\tau'_{a}=O\tau_{a}O^{-1}$:
\begin{eqnarray}
\tau'_{1}=\;\cos \theta \;\cos \phi \;\tau_{1}-
\;\sin \phi \;\tau_{2}\;+\;\sin \theta\;\cos \phi \;\tau_{3}\;,
\nonumber
\\
\tau'_{2}=\cos \theta \;\sin \phi \;\tau_{1}-
\;\cos \phi \;\tau_{2}\;+\;\sin \theta\;\sin \phi \;\tau_{3}\;,
\nonumber
\\
\tau'_{3}=-\;\sin \phi \;\tau_{1}+
\;\cos \theta \;\tau_{3}\;.
\nonumber
\end{eqnarray}

\noindent
Thus, $J'_{a}$ turns out to be
\begin{eqnarray}
J'_{1}=l_{1}+\frac{\cos \phi}{\sin \theta}\;S_{3}\;,
\qquad
J'_{2}=l_{2}+\frac{\sin \phi}{\sin \theta}\;S_{3}\;,
\qquad
J'_{3}=l_{3}\;.
\nonumber
\\
\vec{J\; '}^{2}=\;-\;\frac{1}{\sin \theta}
\partial_{\theta}\;\sin \theta \;\partial_{\theta}\;+\;
\frac{
-\partial^{2}_{\;\phi}\;+\;2\;i\;\partial_{\phi}\;S_{3}\cos \theta\;+
\;S^{2}_{3}}
{\sin^{2}\theta}\;.
\label{1.24}
\end{eqnarray}

For the following  it will be convenient to have  $S_{3}$ diagonal which  can be achieved  through
going to the so-called cyclic basis [2]:
\begin{eqnarray}
 \Psi '' = U \Psi ' \; ,
\qquad  U =
1 \oplus (1 \oplus U_{3}) \oplus (1 \oplus U_{3}) \oplus(U_{3}\oplus U_{3}) \;
\nonumber
\end{eqnarray}
$$
U_{3} = \left |
 \begin{array}{ccc}
- 1 /\sqrt{2}  &  i /  \sqrt{2}  &  0  \\
0  &  0  &  1  \\
1 / \sqrt{2}  &  i  / \sqrt{2}  &  0
\end{array} \right | \; , \;\;
U^{-1} _{3} = U^{+}_{3} = \left |
 \begin{array}{ccc}
- 1 /\sqrt{2}  &  0  & 1 /  \sqrt{2}    \\
-i / \sqrt{2}  &  0  &  -i / \sqrt{2}   \\
0    &  1    &  0
\end{array} \right | \; .
$$

\noindent  It is easily verified
\begin{eqnarray}
U_{3}  \tau_{1} U_{3}^{-1} =
{1 \over \sqrt{2}} \left |  \begin{array}{ccc}
0  &  -i   &  0  \\
-i  &  0  &  -i  \\
0  &  -i  &  0
\end{array} \right | =  \tau'_{1} \; , \;
\nonumber
\\
U_{3}  \tau_{2} U_{3}^{-1} =
{1 \over \sqrt{2}}
\left |  \begin{array}{ccc}
0  &  -1  &  0  \\
1  & 0  &  -1  \\
0  &  1  &  0
\end{array} \right | =  \tau'_{2}\; , \;
\nonumber
\\
U_{3}  \tau_{3} U_{3}^{-1} =
- i \; \left |  \begin{array}{rrr}
+1  &  0  &  0  \\
0  &  0  &  0   \\
0  &  0  &  -1
\end{array} \right | = \tau'_{3} \; . \;
\nonumber
\end{eqnarray}

\noindent
In this cyclic representation  the matrix  $S_{3}$ is diagonal indeed:
$$
S_{3} = \mbox{diag} \; (0 ;\;  0 ,  \; 1,  \; 0,  \; -1;  \;
        0,    \; 1 , \; 0 ,  \; -1 ; \;
        1 , \; 0  , \; -1 ; \;
        1 , \; 0 , \; -1 \;) \; .
$$

\noindent
Also we will be needing  cyclic  quantities (all marks reminding about the use of
spherical tetrad and  cyclic  representation basis will be omitted below)
\begin{eqnarray}
\vec{e}_{1} = ( -{1 \over \sqrt{2}}, \; 0  , \; {1 \over \sqrt{2}} )\; , \;
\vec{e}_{2} = ( -{i \over \sqrt{2}},\;  0  , \; -{i \over \sqrt{2}} )\; , \;
\vec{e}_{3} = ( 0 , 1  , 0)\; , \;
\nonumber
\\
\vec{e}^{\;t} _{1} =
\left | \begin{array}{c}
 - {1 \over \sqrt{2}} \\  0  \\  {1 \over \sqrt{2}}
\end{array} \right | \; , \qquad
\vec{e}^{\;t}_{2} =
\left  | \begin{array}{c}
  {i \over \sqrt{2}} \\  0  \\  {i \over \sqrt{2}}
\end{array} \right | \; , \qquad
\vec{e}^{\;t}_{3} =
\left |  \begin{array}{c}
 0  \\  1  \\  0
\end{array} \right | \; .
\nonumber
\end{eqnarray}

\section{ Separation of variables  for a free particle }

\hspace{5mm}
Now we are  going to construct proper  functions of $\vec{J}^{\;2},J_{3}$-operators.
In accordance  with  the known differential  relations [13]  for Wigner's functions
$ D_{-m,s} ^{j} (\phi, \theta,0) $
\begin{eqnarray}
-i \;   \partial_{\phi} D_{-m, s} ^{j} = m \; D_{-m,s }^{j}  \; ,
\nonumber
\\
(\; -\frac{1}{\sin {\theta}}\partial_{\theta}\sin{\theta}\partial_{\theta}+
\frac{m^{2}+2m\sigma \cos {\theta}+\sigma^{2}}{\sin^{2}{\theta}} \; ) \;
D_{-m, s}^{j}=
j(j+1)D^{j}_{-m, s}  \; ,
\nonumber
\end{eqnarray}

\noindent
the most  general  form of  those proper functions is (see [14])
\begin{eqnarray}
\Psi (x) =
\{  \; C (x) , \; C_{0}(x) ,\; \vec{C}(x) ,\; \Phi_{0} (x) ,
\; \vec{\Phi}(x) , \; \vec{E}(x) , \vec{H}(x) \; \} \; ,
\nonumber
\\
C(x)  = e^{-i\epsilon t} \;  C (r) \;  D_{0} \; , \;\;
C_{0}(x) = e^{-i\epsilon t}   \; C_{0}(r) \;  D_{0}   \; , \;\;
\Phi_{0}(x) =    e^{-i\epsilon t}\; \Phi_{0} (x) \; D_{0} \; ,
\nonumber
\\
\vec{C} (x) =  e^{-i\epsilon t} \;
\left |  \begin{array}{l}
 C_{1}(r)  \; D_{-1}  \\
 C_{2}(3) \; D_{0}    \\
 C_{3}(r) \; D_{+1}      \end{array} \right | \; , \;
\vec{\Phi}(x)   = e^{-i\epsilon t}\;
\left |  \begin{array}{l}
\Phi_{1} (r)\; D_{-1}  \\
\Phi_{2}(r) \; D_{0}   \\
\Phi_{3}(r) \; D_{+1}  \end{array} \right | \; ,
\nonumber
\\
\vec{E} (x)  = e^{-i\epsilon t} \;
\left  | \begin{array}{l}
E_{1}(r) \;  D_{-1} \\
E_{2}(r) \;  D_{0}   \\
E_{3}(r) \;  D_{+1}     \end{array} \right | \; , \;\;
\vec{H} (x) = e^{-i\epsilon t} \;
\left |
 \begin{array}{l}
H_{1}(r)  \; D_{-1}  \\
H_{2}(r)  \; D_{0}   \\
H_{3}(r)  \; D_{+1}    \end{array} \right |\; ,
\label{2.2a}
\end{eqnarray}

\noindent   where the notation is used
$$
D_{s}=
D_{-m,s}^{j}(\phi,\;\theta,\;0) \; , \;\; s = 0,\;+1,\;-1 \; .
$$

\noindent We will need  recurrence relations [13]
\begin{eqnarray}
\partial_{\theta} \;  D_{-1} =   (1/2) \; ( \;
a  \; D_{-2} -    \nu  \; D_{0} \; ) \; ,
\nonumber
\\
\frac {-m+\cos{\theta}}{\sin {\theta}} \; D_{-1} =
(1/2) \; ( \; - a \; D_{-2} -
\nu \; D_{0} \; ) \; ,
\nonumber
\\
\partial_{\theta}  \; D_{0} = (1/2) \; ( \;
\nu  \; D_{-1} - \nu  \; D_{+1} \; ) \; ,
\nonumber
\\
\frac {-m}{\sin {\theta}} \; D_{0} =
(1/2) \; (  \; - \nu \; D_{-1} -  \nu  \; D_{+1} \; ) \; ,
\nonumber
\\
\partial_{\theta} \; D_{+1} =
(1/2) \; ( \; \nu  \; D_{0} - a  \; D_{+2} \; ) \; ,
\nonumber
\\
\frac {-m-\cos{\theta}}{\sin {\theta}}  \; D_{+1} =
(1/2) \; ( \; - \nu \; D_{0} -
\ a  \; D_{+2} \; ) \;  .
\label{2.3}
\end{eqnarray}

\noindent where
$
\nu =  \sqrt{j(j+1)} \; , \;\; a = \sqrt{(j-1)(j+2)} \; .$

Now let us proceed to  separate  variables  in  the main equation
 (\ref{1.10a}). Accounting block-structure of  all quantities involved
after  relevant calculation we  arrive at
\begin{eqnarray}
\partial_{t} \; C_{0} \; -  \;
(\partial_{r} + {2\over r})\; \vec{e}_{3} \; \vec{C} \;  - \;
{1 \over r} \; (\; \vec{e}_{1} \; \partial_{\theta} \;  +  \;
\vec{e}_{2} \; { \partial_{\phi} + \tau_{3} \cos \theta \over \sin \theta }\; ) \;
\vec{C} \; -\; m \; C = 0 \; ,
\nonumber
\\
             -  (\partial_{r} + {2 \over r}) \; \vec{e}_{3} \; \vec{E}  \; -  \;
{1 \over r} \; (\; \vec{e}_{1} \; \partial_{\theta} +
\vec{e}_{2} \; { \partial_{\phi} + \tau_{3} \cos \theta \over \sin \theta }\; ) \;
\vec{E} \; -\;  m \; C_{0} = 0 \; ,
\nonumber
\\
 - \; \partial_{t} \; \vec{E} \; -  \;
(\partial_{r} + {1\over r}) \; \tau_{3} \; \vec{H} \; - \;
{1 \over r} \; (\; \tau_{1} \; \partial_{\theta} +
\tau_{2} \; { \partial_{\phi} + \tau_{3} \cos \theta \over \sin \theta }\; ) \;
\vec{H} \; - \; m \; \vec{C} = 0 \; ,
\nonumber
\\
\sigma \partial_{t} \;  C \;
 - \; (\partial_{r} + {2\over r}) \; \vec{e}_{3}\; \vec{E} \; - \;
{1 \over r} \; (\; \vec{e}_{1} \; \partial_{\theta} +
\vec{e}_{2} \; { \partial_{\phi} + \tau_{3} \cos \theta \over \sin \theta } \;) \;
\vec{E} \; -\;  m \; \Phi_{0} = 0 \; ,
\nonumber
\\
 \partial_{t} \; \vec{E} \; - \;
(\partial_{r} +  {1 \over r}) \; \tau_{3} \; \vec{H} \; + \; \sigma \;
 \vec{e}^{\;\; t} _{3} \partial_{r} C \; +
\nonumber
\\
+ \; \sigma \; {1 \over r} \;
(\; \vec{e}^{\;\;t}_{1} \; \partial_{\theta}   \; +  \;
 \; \vec{e}^{\;\;t}_{2} { \partial_{\phi} \over
\sin \theta } ) \; C  -
{1 \over r} \; (\; \tau_{1} \; \partial_{\theta} +
\tau_{2} \; { \partial_{\phi} + \tau_{3} \cos \theta \over \sin \theta }\; ) \;
\vec{H} \; - \; m \; \vec{\Phi} = 0   \; ,
\nonumber
\\
\partial_{t} \; \vec{\Phi} \; - \; \vec{e}^{\;\;t}_{3} \;
\partial_{r} \Phi_{0} \; - \;
\; {1 \over r} \;
(\; \vec{e}^{\;\;t}_{1} \; \partial_{\theta}   \; +  \;
 \; \vec{e}^{\;\;t}_{2} \; { \partial_{\phi} \over
\sin \theta } \; ) \; \Phi _{0} \;  - \;  m  \; \vec{E} = 0 \; ,
\nonumber
\\
( \partial_{r} + {1 \over r}  ) \tau_{3}  \; \vec{\Phi} \; + \;
{1 \over r} \; ( \; \tau_{1} \; \partial_{\theta} +
\tau_{2} \; { \partial_{\phi} + \tau_{3} \cos \theta \over \sin \theta } \; ) \;
\vec{\Phi} \; - \; m \; \vec{H} = 0 \; .
\label{2.4}
\end{eqnarray}

Further with the use of intermediate results
\begin{eqnarray}
(\; \vec{e}_{1} \; \partial_{\theta} \;  +  \;
\vec{e}_{2} \; { \partial_{\phi} + \tau_{3} \cos \theta \over \sin \theta } \;  ) \;
\vec{C} (x)\; =
= e^{-i\epsilon t} \; \left \{ \; \vec{e}_{1} \;
\left |  \begin{array}{c}
C_{1} \; \partial_{\theta} D_{-1} \\
C_{2} \; \partial_{\theta} D_{\;\;0} \\
C_{3} \; \partial_{\theta} D_{+1}
\end{array} \right |    \; + \right.
\nonumber
\\
\left. + \;   (-i)  \; \vec{e}_{2} \;
\left |  \begin{array}{c}
C_{1}  \; { -m + \cos \theta \over \sin \theta } \;D_{-1} \\
C_{2}  \; { -m \over \sin \theta }  \;\;D_{\;\;0} \\
C_{3}  \;  { -m + \cos \theta \over \sin \theta }  \; D_{+1}
\end{array} \right | \; \right  \} =
 e^{-i\epsilon t} \; {\nu \over \sqrt{2}} \; (C_{1} + C_{3}) \;   D_{0} \; ,
\nonumber
\\
(\; \tau_{1} \; \partial_{\theta} +
\tau_{2} \; { \partial_{\phi} + \tau_{3} \cos \theta \over \sin \theta }\; ) \;
\vec{H} (x) =
e^{-i\epsilon t} \;  \left \{ \; \tau_{1}
\left |  \begin{array}{c}
H_{1} \; \partial_{\theta} D_{-1} \\
H_{2} \; \partial_{\theta} D_{0} \\
H_{3} \; \partial_{\theta} D_{+1}
\end{array}  \right |
\; + \right.
\nonumber
\\
\left. +  \; \tau_{2} \; \left |  \begin{array}{c}
H_{1}  \; { -m + \cos \theta \over \sin \theta } \;D_{-1} \\
H_{2}  \; { -m \over \sin \theta }  \;\;D_{\;\;0} \\
H_{3}  \;  { -m + \cos \theta \over \sin \theta }  \; D_{+1}
\end{array} \right |  \right  \} =
e^{-i \epsilon t} \; {i \nu \over \sqrt{2}} \; \left |  \begin{array}{c}
- H_{2} \; D_{-1} \\ (H_{1} - H_{3} ) \; D_{0} \\
H_{2} D_{+1}  \end{array} \right | \; ,
\nonumber
\\
(\; \vec{e}^{\;\;t}_{1} \; \partial_{\theta}   \; +  \;
 \; \vec{e}^{\;\;t}_{2} { \partial_{\phi} \over
\sin \theta }\; ) \; C (x) =
- e^{-i \epsilon t }  \;  {\nu \over \sqrt{2}} \; \left | \begin{array}{c}
C \; D_{-1} \\  0  \\ C \; D_{+1}  \end{array} \right | \; ,
\nonumber
\end{eqnarray}

\noindent
we get to a radial equation system
(the notation   $ \nu = \sqrt{j(j+1)} / \sqrt{2}$ is used)
\begin{eqnarray}
-i\; \epsilon \; C_{0}-\;\;({d \over d r} + \frac {2}{r})\; C_{2} -
\frac{\nu}{r } \; (C_{1}+C_{3}) = m \;  C \; .
\label{2.6a}
\end{eqnarray}
\begin{eqnarray}
- \; ({d \over d r} + \frac {2}{r}) \; E_{2}
- \; \frac{\nu}{r}\; (E_{1}+E_{3}) = m \; C_{0} \;    ,
\nonumber
\\
+i \; \epsilon \; E_{1} + i ({d \over d r} + \frac {1}{r}) \; H_{1} \; + \;
i \frac{\nu}{r}\; H_{2}  = m \; C_{1}  \;   ,
\nonumber
\\
+i \; \epsilon \; E_{2} \;  -
i \; \frac{\nu}{r}\; (H_{1} - H_{3}) = m \; C_{2}  \;   ,
\nonumber
\\
+i \; \epsilon \; E_{3}  - i ({d \over d r} + \frac {1}{r}) \; H_{3} -
i \frac{\nu}{r} \; H_{2} = m \; C_{3}  \;  .
\label{2.6b}
\end{eqnarray}
\begin{eqnarray}
- i  \; \epsilon \; \sigma C\;\; - \;\;({d \over d r}  + \frac {2}{r}) \; E_{2} -
\frac{\nu}{r}\;  (E_{1} + E_{3}) = m  \; \Phi_{0}  \; ,
\nonumber
\\
+ i \; \epsilon \; E_{1} \;\; + i \;({d \over d r} + \frac {1}{r}) \; H_{1} +
i \frac{\nu}{r} \; H_{2} - \sigma \frac{\nu}{r }  \; C
=  m \Phi_{1} \; ,
\nonumber
\\
+ i \; \epsilon \; E_{2} \;\; + \;\sigma\; {d \over d r} \;C -
i\frac{\nu}{r}\; (H_{1} - H_{3})
=  m \;  \Phi_{2}  \; ,
\nonumber
\\
+i \; \epsilon \; E_{3} \;\; - i \;({d\over d r} + \frac {1}{r})\; H_{3} -
i \frac{\nu}{r} \; H_{2} - \; \sigma \; \frac{\nu}{r} \; C
=  m \; \Phi_{3}  \;   .
\label{2.6c}
\end{eqnarray}
\begin{eqnarray}
-i \; \epsilon \; \Phi_{1}\;\;  +
\frac{\nu}{r} \; \Phi_{0} =  m \;  E_{1}  \;     ,
\nonumber
\\
-i \; \epsilon \; \Phi_{2} \;\; - \;{d \over  d r} \; \Phi_{0}
=  m \; E_{2}  \;    ,
\nonumber
\\
-i \; \epsilon \; \Phi_{3} \;\;  +
\frac{\nu}{r} \; \Phi_{0}  = m \; E_{3} \;   ,
\nonumber
\\
 - i\; ({d \over d r} + \frac {1}{r}) \; \Phi_{1} -
i\frac{\nu}{r} \; \Phi_{2} =  m \; H_{1} \;  ,
\nonumber
\\
+i\frac{\nu}{r}\; (\Phi_{1} - \Phi_{3}) =  m \; H_{2} \; ,
\nonumber
\\
\qquad \qquad +i\;({d \over d r} + \frac {1}{r})\; \Phi_{3} +
i\frac{\nu}{r} \Phi_{2} =  m \;  H_{3} \; .
\label{2.6d}
\end{eqnarray}

Concurrently we will diagonalize $P$-inversion operator. In Cartesian  tetrad it ha s a
conventional form
\begin{eqnarray}
\Pi =  [\; 1 \oplus (1 \oplus -I) \oplus(1 \oplus -I)\oplus (-I \oplus +I) \; ]\;
 \hat{P} \; , \;\;\hat{P} \Psi (\vec{r}) = \Psi (-\vec{r}) \; .
\label{2.7}
\end{eqnarray}

\noindent
which after translating  to the spherical-cyclic basis will become
\begin{eqnarray}
\hat{\Pi} '  =  [ 1 \oplus (1 \oplus \Pi_{3}) \oplus ( 1 \oplus \Pi_{3}) \oplus (\Pi_{3} \oplus
-\Pi_{3})\;]\;
  \hat{P} \; , \;\;
\Pi_{3}  =    \left |  \begin{array}{rrr}
0  &  0  &  -1  \\
0  &  -1 &   0  \\
-1  &  0  &  0   \end{array} \right |\; .
\label{2.8}
\end{eqnarray}

The eigenvalue equation  $\hat{\Pi} \Psi = P \; \Psi $ , with the use of
$ \hat{P} D_{s} = (-1)^{j} \; D_{-s}$,  supplies us with solutions of two sorts:
\begin{eqnarray}
\underline{P=(-1)^{j+1}} \; ,
\qquad
C = 0 \; , \;\; C_{0} = 0 \; , \;\; C_{3} = - C_{1} \; , \;\; C_{2} = 0 \; ,
\nonumber
\\
\Phi_{0} = 0 \; , \qquad \Phi_{3} = - \Phi_{1} \; , \qquad \Phi_{2} = 0 \; ,
\nonumber
\\
E_{3} = - E_{1} \; , \qquad  E_{2} = 0\;  ,\qquad H_{3}  = H_{1} \;\; ;
\label{2.10}
\end{eqnarray}
\begin{eqnarray}
\underline{P=(-1)^{j}} , \qquad \qquad \qquad
C_{3} = + C_{1}  \; , \qquad  \Phi_{3} = + \Phi_{1} \; , \;\;
\nonumber
\\
\qquad  \qquad
E_{3} = + E_{1} \; , \; H_{3} = - H_{1} \; , \; \;
H_{2} = 0 \; .
\label{2.11}
\end{eqnarray}

It is  easily verified that both (\ref{2.10}) and (\ref{2.11})  are consistent with the above radial system (2.6);
at this  we will get two sub-systems respectively:
\begin{eqnarray}
\underline{P=(-1)^{j+1}} ,
\qquad
+i \; \epsilon  \; E_{1} +  i \; ({d \over d r} \; + \; { 1 \over r}) \; H_{1} \; + \;
i\;  {\nu \over r} H_{2}
= m\;  C_{1} \; , \;\;
\nonumber
\\
+i \; \epsilon \; E_{1} \; + \;  i \; ({d \over d r} \; + \; { 1 \over r}) \; H_{1} + \;
 i \; {\nu \over r} \; H_{2}
= m \; \Phi _{1} \; ,
\;\;
-i \; \epsilon \; \Phi_{1}  = m \; E_{1} \; , \;\;
\nonumber
\\
-i \; ({d \over d r} \; +\;  {1 \over r}) \; \Phi_{1}   = m \; H_{1} \; , \;\;
2i \; { \nu \over r } \; \Phi_{1} = m \; H_{2}  \; ;
\label{2.12}
\end{eqnarray}
\begin{eqnarray}
\underline{P=(-1)^{j}} , \qquad
-i \; \epsilon \; C_{0}\;  - \; ( {d \over d r} \;+ \;{2 \over r})\; C_{2} \;-\;
2\; {\nu \over r}\; C_{1} = m \; C \; ,
\nonumber
\\
- ({ d \over d r} + {2 \over r}) \; E_{2} \; - \; 2 \; {\nu \over r} \; E_{1} = m \; C_{0} \; ,
\nonumber
\\
+ i \; \epsilon \; E_{1} \; +\; i \; ( {d \over d r} \; + \; {1 \over r}) \; H_{1}
= m \; C_{1} \; ,
\nonumber
\\
+\;  i \; \epsilon \;  E_{2} \; - \; 2i \; {\nu \over r} \; H_{1} = m \; C_{2} \; ,
\nonumber
\\
- i \; \epsilon \; \sigma\; C - ({d \over d r} \; + \; {2\over r}) \; E_{2} \; -\;
2 {\nu \over r} \; E_{1} = m \; \Phi_{0}\; ,
\nonumber
\\
+ i \; \epsilon \; E_{1} \; + \;i ( {d \over d r} + {1 \over r}) \; H_{1} \;- \;\sigma\; {\nu \over r} \; C =
m \; \Phi_{1} \; ,
\nonumber
\\
+ i \; \epsilon E_{2} \; -2i  \; {\nu \over r} \;  H_{1} \; +\; \sigma\; {d \over d r} \; C =
m \; \Phi_{2} \; ,
\nonumber
\\
-i\; \epsilon \; \Phi_{1} \; + \; {\nu \over r} \; \Phi_{0} = m \; E_{1} \; ,
\nonumber
\\
-i \;\epsilon \; \Phi_{2}  \; - \; {d \over d r} \; \Phi_{0} = m \; E_{2} \; ,
\nonumber
\\
-i \; ({d \over d r} \; + \; {1\over r}) \; \Phi_{1} \; -
\; i {\nu \over r} \;  \Phi_{2} = m \; H_{1} \; .
\label{2.13}
\end{eqnarray}

\section{ Vector particle in Coulomb field }

\hspace{5mm}
In presence  of external electromagnetic fields the original equation (\ref{1.3})
 will become
\begin{eqnarray}
[ \; \Gamma ^{\alpha }(x)\; ( \partial_{\alpha} \;  +  \;
B_{\alpha }(x) - i {e \over \hbar c} A_{\alpha} ) \; - {m c  \over \hbar } \;  ] \;\Psi  (x)  = 0 \; .
\label{3.2}
\end{eqnarray}

\noindent
As Coulomb field is described by the  potential
$
A_{\alpha}(x) = ( z e /  r , 0 , 0 , 0  ) \; ,
$
then most of the calculation in  previous Sec. 2 and Sec. 3  need not  to be  repeated -- you
can immediately turn to radial equations  with  one  formal change ( $ze^{2} = \alpha$)
$$
\epsilon \qquad  \Longrightarrow \qquad  ( \epsilon + {ze^{2} \over r})  \; .
$$

\noindent Thus,  we  will obtain
\begin{eqnarray}
\underline{P=(-1)^{j+1}} ,\qquad
+i (\epsilon + {\alpha \over r}) \;  E_{1} \; + \; i ({d \over d r} +{ 1 \over r}) \;
H_{1} \; + \; i {\nu \over r} \; H_{2}
= m \;  C_{1} \; ,
\nonumber
\\
+ i  (\epsilon + {\alpha \over r}) \; E_{1}  \;+  \; i ({d \over d r} +{ 1 \over r}) \; H_{1} \; + \;
 i\; {\nu \over r} \; H_{2}
= m \; \Phi _{1} \; ,
\nonumber
\\
-i \; (\epsilon + {\alpha \over r}) \;  \Phi_{1}  = m \; E_{1} \; ,
\;\;
-i  \;({d \over d r} + {1 \over r}) \; \Phi_{1}   = m \; H_{1} \; ,
\;\;2i { \nu \over r } \; \Phi_{1} = m \; H_{2}  \; ;
\label{3.4}
\end{eqnarray}
\begin{eqnarray}
\underline{P=(-1)^{j}} , \qquad
-i ( \epsilon  + {\alpha \over r}) \; C_{0} \; - \; ( {d \over d r} + {2 \over r}) \; C_{2} \; -
\; 2 {\nu \over r} \; C_{1} = m \; C \; ,
\nonumber
\\
- ({ d \over d r} + {2 \over r}) \; E_{2} \;  - \; 2 {\nu \over r} \; E_{1} = m \; C_{0} \; ,
\nonumber
\\
+ i (\epsilon  + {\alpha \over r}) \; E_{1} \; + \; i ( {d \over d r} + {1 \over r}) \; H_{1} = m \; C_{1} \; ,
\nonumber
\\
+ i ( \epsilon + {\alpha \over r}) \; E_{2} \; - \; 2i {\nu \over r} \; H_{1} = m \; C_{2} \; ,
\nonumber
\end{eqnarray}
\begin{eqnarray}
-i (\epsilon  + {\alpha \over r}) \; \sigma\; C \;  - \; ({d \over d r}  + {2\over r}) \;  E_{2} \; -
\; 2 {\nu \over r} \; E_{1} = m \;\Phi_{0}\; ,
\nonumber
\\
+ i (\epsilon + {\alpha \over r}) \; E_{1} \; + \; i ( {d \over d r} + {1 \over r}) \;H_{1} \;- \;
\sigma\; {\nu \over r}\; C = m \; \Phi_{1} \; ,
\nonumber
\\
+ i (\epsilon  + {\alpha \over r}) \;  E_{2} \; - \; 2i {\nu \over r} \;  H_{1} \; + \; \sigma\;
{d \over d r} \; C = m\;  \Phi_{2} \; ,
\nonumber
\\
-i (\epsilon + {\alpha \over r}) \;  \Phi_{1} \; + \; {\nu \over r} \; \Phi_{0} = m \; E_{1} \; , \;
\nonumber
\\
-i (\epsilon + {\alpha \over r})\; \Phi_{2} \;  - \; {d \over d r} \; \Phi_{0} = m \; E_{2} \; ,
\nonumber
\\
-i ({d \over d r} +{1\over r}) \; \Phi_{1} \; - \; i {\nu \over r} \;  \Phi_{2} = m \; H_{1} \; .
\label{3.5}
\end{eqnarray}

Concerning eqs.  (\ref{3.4}) for $P = (-1)^{J+1}$ not much need to be  done  to arrive  at a final
second-order  differential  relation. Actually,  from (\ref{3.4}) it follows
\begin{eqnarray}
C_{1} (r) = \Phi_{1}(r) \; ,
\qquad
m \; E_{1} =  - i \; ( \epsilon + {\alpha \over r}) \;  \Phi_{1} \; ,
\nonumber
\\
m \; H_{1} =  -i \; ({d \over d r} + {1 \over  r}) \;
\Phi_{1} \; , \; \;
 m \; H_{2} = 2 i \; { \nu \over r} \;  \Phi_{1} \;
\label{3.6a}
\end{eqnarray}

\noindent and for $\Phi_{1}$
\begin{eqnarray}
\left [ \;{ d^{2} \over dr^{2} } + {2 \over r} {d \over d r}  + (\epsilon + {\alpha \over r})^{2} -
{j(j+1) \over r^{2}} \; \right ] \; \Phi_{1} = 0 \;
\label{3.6b}
\end{eqnarray}

\noindent
what is  the common equation  in the theory of a usual scalar  particle in  Coulomb field.
Its solutions are well-known.

\begin{quotation}

So, in part,  the extended  theory under consideration behaves like an ordinary  scalar particle in the
Coulomb field. Remaining case when $P=(-1)^{j+1}$  turns out to be  much more involved.
For instance, even the simplest case $j=0$  leads us to an equation too
difficult for analytical treatment.

\end{quotation}

Now let us turn to  (\ref{3.5}).
Taking $C_{0}, C_{1},C_{2}$ from the first equation for $m^{2}C(r)$ it follows
\begin{eqnarray}
m^{2}\; C = -i (\epsilon + {\alpha \over r} ) \;
\left [ \; - ( {d \over d r} + {2 \over r} ) \; E_{2} \; -\;
2 {\nu \over r} \; E_{1} \; \right ] \;-
\nonumber
\\
-\; ( {d \over d r} + {2 \over r} ) \; \left [
 \; i ( \epsilon + {\alpha \over r}) \; E_{2} \;- \;
2i{ \nu \over r} \;H_{1} \; \right  ]\; -
\; 2 \;{\nu \over r} \;\left  [ \; i ( \epsilon + {\alpha \over r})\; E_{1} \;+\; i ( {d \over d r} + {1 \over r})\;
 H_{1}\; \right ]  \; ,
\nonumber
\end{eqnarray}

\noindent that is
\begin{eqnarray}
 C (r) =
+\; {i \alpha \over  m^{2}\; r^{2}} \; E_{2} (r) \; .
\label{3.7a}
\end{eqnarray}

Substitution this expression into remaining equation in (\ref{3.5}), we  get  to
\begin{eqnarray}
{\sigma \;  \alpha \over m^{2}\; r^{2} } \; (\epsilon + {\alpha \over r}) \; E_{2} \; - \;
 ({d \over d r} +{2 \over r}) \;  E_{2} \; - \; {2 \nu\over r} \; E_{1} = m \; \Phi_{0}\; ,
\nonumber
\\
i(\epsilon + {\alpha \over r}) \;   E_{1} \; + \;i ( {d \over d r} \;+ \; {1 \over r})
\; H_{1} \;
-\;
 {i \sigma  \alpha  \nu \over m^{2} r^{3}} \;  E_{2} =
m\;  \Phi_{1} \; ,
\nonumber
\\
i(\epsilon  + {\alpha \over r})  \; E_{2} \; - \; 2 {i\nu \over r} \; H_{1} \; + \;
i \sigma\; {d \over d r}
\; ({\alpha \over m^{2}\;r^{2}} \;  E_{2})  =
m \; \Phi_{2} \; ,
\nonumber
\end{eqnarray}
\begin{eqnarray}
-i(\epsilon + {\alpha \over r}) \; \Phi_{1} + {\nu \over r}\; \Phi_{0} = m \; E_{1} \; ,
\nonumber
\\
-(\epsilon + {\alpha \over r}) \; \Phi_{2}  \; - \;  {d \over d r}  \; \Phi_{0} = m \; E_{2} \; ,
\nonumber\\
-i ({d \over d r} + {1 \over r} ) \; \Phi_{1} \;  -i  \;  {\nu \over r}  \; \Phi_{2} =
m  \; H_{1} \; .
\label{3.7b}
\end{eqnarray}

\noindent
Unfortunately,  we have not been able to  proceed  with these  equations successfully.

For instance, let us consider in some detail the  simplest case of
$j=0$. For this case a special initial form of the wave function must be  used (compare it
 with (\ref{2.2a} )
\begin{eqnarray}
\Psi_{j=0} (x) =
\{  \; C (x) , \; C_{0}(x) ,\; \vec{C}(x) ,\; \Phi_{0} (x) ,
\; \vec{\Phi}(x) , \; \vec{E}(x) , \vec{H}(x) \; \} \; ,
\nonumber
\\
C(x)  = e^{-i\epsilon t} \;  C (r) \;  D_{0} \; , \;\;
C_{0}(x) = e^{-i\epsilon t}   \; C_{0}(r) \;  D_{0}   \; , \;\;
\Phi_{0}(x) =    e^{-i\epsilon t}\; \Phi_{0} (x) \; D_{0} \; ,
\nonumber
\\
\vec{C} (x) =  e^{-i\epsilon t} \;
\left |  \begin{array}{c}
 0     \\
 C_{2}(3)     \\
 0       \end{array} \right | \; , \;
\vec{\Phi}(x)   = e^{-i\epsilon t}\;
\left | \begin{array}{c}
0   \\
\Phi_{2}(r)    \\
0   \end{array} \right | \; ,
\nonumber
\\
\vec{E} (x)  = e^{-i\epsilon t} \;
\left | \begin{array}{c}
0  \\
E_{2}(r)   \\
0      \end{array} \right | \; , \;\;
\vec{H} (x) = e^{-i\epsilon t} \;
\left | \begin{array}{c}
0   \\
H_{2}(r)     \\
0      \end{array} \right |\; .
\label{3.7}
\end{eqnarray}

\noindent
Operator $\Sigma _{\theta \phi }$ being applied
to that functions vanishes identically; $P$-operator takes on such solutions the proper
 value $-1$. So setting $\nu = 0$ in (\ref{3.7b})  and taking in mind eqs. (\ref{3.7})
 we  get only three nontrivial equations:
\begin{eqnarray}
{\sigma \;  \alpha \over m^{2}\; r^{2} } \; (\epsilon + {\alpha \over r}) \; E_{2} \; - \;
 ({d \over d r} +{2 \over r}) \;  E_{2} = m \; \Phi_{0}\; ,
\qquad 0 =0 \; ,
\nonumber
\\
i(\epsilon  + {\alpha \over r})  \; E_{2} \;  + \;
i \sigma\; {d \over d r}
\; ({\alpha \over m^{2}\;r^{2}} \;  E_{2})  =
m \; \Phi_{2} \; ,
\qquad 0 =0 \; ,
\nonumber
\\
-(\epsilon + {\alpha \over r}) \; \Phi_{2}  \; - \;  {d \over d r}  \; \Phi_{0} = m \; E_{2} \; ,
\qquad
0= 0 \; .
\label{3.7c}
\end{eqnarray}

\noindent
Excluding variables $\Phi_{0},\Phi_{2}$, we arrive at an  equation for $E_{2}$:
\begin{eqnarray}
m^{2} \; E_{2} = - ( \epsilon + {\alpha \over r}) \left [
i(\epsilon  + {\alpha \over r})  \;   + \;
i \sigma\; {d \over d r}
\; {\alpha \over m^{2}\;r^{2}}
\right ]\; \;  E_{2} \; -
\nonumber
\\
- {d \over dr} \; \left [
{\sigma \;  \alpha \over m^{2}\; r^{2} } \; (\epsilon + {\alpha \over r}) \; E_{2} \; - \;
 ({d \over d r} +{2 \over r}) \;  E_{2} \right ] \; ,
\nonumber
\end{eqnarray}

\noindent from which it follows    a 2-oder differential equation
\begin{eqnarray}
\left [
{d^{2} \over dr^{2}} + {2 \over r}\; {d\over dr}
 +
(\epsilon +{\alpha \over r })^{2} - m^{2} - {2 \over r^{2}} + {\sigma \; \alpha^{2} \over m^{2} r^{4}}
\right ]\; E_{2}(r) =0 \; .
\label{3.9}
\end{eqnarray}

\section{Particle in magnetic monopole field}

\hspace{5mm}
Now we will consider the vector particle in  external Dirac monopole field [15]. Initial
wave equation  is
\begin{eqnarray}
[ \; \Gamma ^{\alpha }(x)\; ( \partial_{\alpha} \;  +  \;
B_{\alpha } - i {e \over \hbar c} A_{\alpha} ) \; - {m c  \over \hbar } \;  ] \;\Psi  (x)  = 0 \; .
\nonumber
\end{eqnarray}

\noindent To describe monopole field  the   Schwinger monopole  potential will be used; the latter after translating to  spherical
coordinates will have  the explicit form
$A_{\alpha} = (0, 0,0, A_{\phi} = g \cos \theta ) \;
$;
$g$ stands for a  magnetic charge.
According to  this  you  are to make  one substitution
\begin{eqnarray}
\partial_{\phi} \Longrightarrow (\partial_{\phi} + i k \; \cos \theta ) \; , \;\;\; \mbox{where}
\; \;\; k = {eg \over \hbar c} \; .
\nonumber
\end{eqnarray}

\noindent Correspondingly, the main equation
can be read as
\begin{eqnarray}
\left [\; \Gamma^{0} \partial_{0}\;  + \; \Gamma^{3}\partial_{r} +
\frac{\Gamma^{1} J^{31} + \Gamma^{2} J^{32} } {r}  \; + \;
\frac{1}{r}\; \Sigma^{k}_{\theta,\phi } \; - \; m \; \right  ] \; \Psi = 0 \; ,
\label{4.3a}
\end{eqnarray}

\noindent where
\begin{eqnarray}
\Sigma^{k}_{\theta,\phi} = \Gamma^{1} \; \partial_{\theta}  +  \Gamma^{2}
{  \partial_{\phi} + ( J^{12} + i k ) \cos \theta   \over \sin \theta } \; .
\label{4.3b}
\end{eqnarray}

\noindent
You may  immediately pass to equations of the type  (\ref{2.4}):
\begin{eqnarray}
\partial_{t} \;C_{0} \; -  \;
(\; \partial_{r} + {2\over r}\; )\; \vec{e}_{3} \; \vec{C} \;  - \;
{1 \over r} \; ( \; \vec{e}_{1} \; \partial_{\theta} \;  +  \;
\vec{e}_{2} \; { \partial_{\phi} + (\tau_{3} + i k) \cos \theta \over \sin \theta } \; ) \;
\vec{C} = m \; C  ,
\nonumber
\\
 -  (\; \partial_{r} + {2 \over r}\; ) \;\vec{e}_{3} \; \vec{E}  \; -  \;
{1 \over r} \; ( \; \vec{e}_{1} \; \partial_{\theta} +
\vec{e}_{2} \; { \partial_{\phi} + (\tau_{3} +i k) \cos \theta \over \sin \theta } \; ) \;
\vec{E} =  m \; C_{0} ,
\nonumber
\\
 - \; \partial_{t}\; \vec{E} \; -  \;
(\; \partial_{r} + {1\over r}\; ) \; \tau_{3} \; \vec{H} \; - \;
{1 \over r} \; ( \; \tau_{1} \; \partial_{\theta} +
\tau_{2} \; { \partial_{\phi} + (\tau_{3}  + i k) \cos \theta \over \sin \theta } \; ) \;
\vec{H} = m \; \vec{C} ,
\nonumber
\\
\sigma \; \partial_{t} \;  C
 -  ( \; \partial_{r} + {2\over r} \; ) \; \vec{e}_{3} \; \vec{E} \; - \;
{1 \over r} \; ( \; \vec{e}_{1} \; \partial_{\theta} +
\vec{e}_{2} \; { \partial_{\phi} + (\tau_{3} + i k ) \cos \theta \over \sin \theta } \; ) \;
\vec{E} =  m \; \Phi_{0}  ,
\nonumber
\\
- \partial_{t} \; \vec{E} \; - \;
(\; \partial_{r} +  {1 \over r} \; ) \; \tau_{3} \; \vec{H} \; + \; \sigma \;
 \vec{e}^{\;t}_{3} \; \partial_{r} C \; +
\; \sigma \; {1 \over r} \;
( \; \vec{e}^{\;t}_{1} \; \partial_{\theta}   \; +  \;
 \; \vec{e}^{\;t}_{2} { \partial_{\phi}  + i k \cos \theta \over
\sin \theta } \; ) \; C  -
\nonumber
\\
- \;
{1 \over r} \; ( \; \tau_{1} \; \partial_{\theta} +
\tau_{2} \; { \partial_{\phi} + (\tau_{3} + i k ) \cos \theta \over \sin \theta } \; ) \;
\vec{H} =  m \; \vec{\Phi}  ,
\nonumber
\\
\partial_{t} \; \vec{\Phi } \; - \; \vec{e}^{\;t}_{3} \;
\partial_{r} \; \Phi_{0} \; - \;
\; {1 \over r} \;
(\; \vec{e}^{\;t}_{1} \; \partial_{\theta}   \; +  \;
 \; \vec{e}^{\;t}_{2} \; { \partial_{\phi}  + i k \cos \theta \over
\sin \theta } \; ) \; \Phi _{0} =  m  \; \vec{E} ,
\nonumber
\\
(\; \partial_{r} + {1 \over r } \; )\;  \tau_{3} \; \vec{\Phi} \; + \;
{1 \over r} \; (\; \tau_{1} \; \partial_{\theta} +
\tau_{2} \; { \partial_{\phi} + (\tau_{3}  + i k ) \cos \theta \over \sin \theta } \;  ) \;
\vec{\Phi } =  m \; \vec{H}  .
\label{4.4}
\end{eqnarray}

It is the point to establish  a suitable form of the  wave  function being  proper one of
rotation symmetry-based  operators. In presence  monopole background
there exist three relevant ones:
\begin{eqnarray}
J^{(k)}_{1}=l_{1}+\frac{\cos \phi}{\sin \theta}\;( S_{3} - k )\;,
\;\;
J^{(k)}_{2}=l_{2}+\frac{\sin \phi}{\sin \theta}\;(S_{3} - k)\;,
\;\;
J^{(k)}_{3}=l_{3}\;.
\label{4.5}
\end{eqnarray}

\noindent
Therefore, two operators to be diagonalized are
\begin{eqnarray}
J^{(k)}_{3}=l_{3}\; , \qquad
\vec{J}_{(k)}^{\;\;2}=\;-\;\frac{1}{\sin \theta}
\partial_{\theta}\;\sin \theta \;\partial_{\theta}\;+\;
\nonumber
\\
+\frac{
-\partial^{2}_{\;\phi}\;+\;2\;i\;\partial_{\phi}\; (S_{3} - k) \cos \theta\;+
\;(S - k)^{2}_{3}}
{\sin^{2}\theta}\;.
\label{4.6}
\end{eqnarray}

\noindent So, the most general form of 15-component wave  function with quantum numbers
$\epsilon , j , m$ will look as
\begin{eqnarray}
\Psi (x) =
\{  \; C (x) , \; C_{0}(x) ,\; \vec{C}(x) ,\; \Phi_{0} (x) ,
\; \vec{\Phi}(x) , \; \vec{E}(x) ,  \vec{H}(x) \; \} \; ,
\nonumber
\end{eqnarray}
\begin{eqnarray}
C(x)  = e^{-i\epsilon t} \;  C (r) \;  D_{k} \; , \;\;\;\;
C_{0}(x) = e^{-i\epsilon t}   \; C_{0}(r) \;  D_{k}   \; , \;\;
\Phi_{0}(x) =    e^{-i\epsilon t}\; \Phi_{0} (x) \; D_{k} \; , \;\;
\nonumber
\\
\vec{C} (x) =  e^{-i\epsilon t} \;
\left | \begin{array}{c}
C_{1}(r)  \; D_{k-1} \\  C_{2}(r) \; D_{k}  \\ C_{3}(r) \; D_{k+1}
\end{array} \right | \; , \;
\vec{\Phi}(x)   = e^{-i\epsilon t}\;
\left | \begin{array}{c}
\Phi_{1} (r)\; D_{k-1} \\ \Phi_{2}(r) \;  D_{k} \\ \Phi_{3} (r) \; D_{k+1}
\end{array} \right | \; , \;
\nonumber
\\
\vec{E} (x)  = e^{-i\epsilon t} \;
\left | \begin{array}{c}
E_{1}(r) \;  D_{k-1}  \\ E_{2}(r) \;  D_{k} \\  \; E_{3}(r) \; D_{k+1}
\end{array} \right | \; , \;
\vec{H} (x) = e^{-i\epsilon t} \;
\left | \begin{array}{c}
H_{1}(r) \;  D_{k-1}  \\  H_{2}(r) \;  D_{k}  \\ H_{3}(r) \;  D_{k+1}
\end{array} \right | \; .
\end{eqnarray}

\noindent   where
$
D_{s}=
D_{-m,s}^{j}(\phi,\;\theta,\;0) \; , \;\; s = k,\;k+1,\;k-1 \; .
$

Now  you should substitute the wave function into  (\ref{4.4}) and exclude variable $t,\theta, \phi$.
At this you will need some known recurrence relations [13]
\begin{eqnarray}
\partial_{\theta } \; D_{\kappa -1} \; = \;
(a \; D_{\kappa-2} - c \; D_{\kappa } )\; ,
\nonumber
\\
{ -m-(\kappa -1) \cos \theta  \over \sin \theta } \; D_{\kappa -1} =
(-a \; D_{\kappa -2} - c \;  D_{\kappa }) \; ,
\nonumber
\\
\partial _{\theta }\; D_{\kappa } \;  = \;
(c \; D_{\kappa-1} -  d \; D_{\kappa +1})\; ,
\nonumber
\\
{- m  - \kappa  \cos \theta \over \sin \theta } \; D_{\kappa } =
(-c \;  D_{\kappa-1} - d \;  D_{\kappa +1})\; ,
\nonumber
\\
\partial _{\theta } \; D_{\kappa +1} \; = \;
(d \; D_{\kappa } - b \;  D_{\kappa +2})\; ,
\nonumber
\\
{-m-(\kappa +1)\cos \theta \over \sin \theta } \;  D_{\kappa +1} \; = \;
(-d \;  D_{\kappa } - b \; D_{\kappa +2})  \; ,
\label{4.11a}
\end{eqnarray}

\noindent where
\begin{eqnarray}
a = {1 \over 2}  \sqrt{(j + \kappa  -1)(j - \kappa  + 2)}\;  , \qquad
b = {1 \over 2}  \sqrt{(j - \kappa  -1)(j + \kappa  + 2)} \; ,
\nonumber
\\
c = {1 \over 2}  \sqrt{(j + \kappa )(j - \kappa + 1)} \; , \qquad
d = {1 \over 2}  \sqrt{(j - \kappa )(j + \kappa  +1)} \; .
\nonumber
\end{eqnarray}

\noindent
Eqs.  (\ref{4.4}) rewritten as
\begin{eqnarray}
\partial_{t} \;  C_{0} \; -  \;
(\partial_{r} + {2\over r})\; \vec{e}_{3} \;  \vec{C} \;  - \;
{1 \over r} \; [ \;  \vec{e}_{1} \; \partial_{\theta} \;  +  \;
\vec{e}_{2} (-i) \; { - m -  (k - s_{3}) \cos \theta \over \sin \theta } \;  ]
  \; \vec{C} =  m \; C \; ,
\nonumber
\\
             -  (\partial_{r} + {2 \over r}) \;\vec{e}_{3} \; \vec{E}  \; -  \;
{1 \over r} \; [ \vec{e}_{1} \; \partial_{\theta} +
\vec{e}_{2} (-i) \; { - m  - (k - s_{3}) \cos \theta \over \sin \theta } ] \;
\vec{E} =   m \; C_{0} \; ,
\nonumber
\\
 - \; \partial_{t} \vec{E} \; -  \;
(\partial_{r} + {1\over r}) \; \tau_{3}  \;\vec{H} \; - \;
{1 \over r} \; [ \tau_{1} \; \partial_{\theta} +
\tau_{2} (-i) \; { -m  -  (k - s_{3} ) \cos \theta \over \sin \theta } ] \;
\vec{H} =  m \;\vec{C}  \; ,
\nonumber
\\
\sigma \partial_{t} C
 -  (\partial_{r} + {2\over r}) \; \vec{e}_{3} \; \vec{E} \; - \;
{1 \over r} \; [ \;  \vec{e}_{1} \; \partial_{\theta} \; +
\vec{e}_{2} (-i) \; { -m  -  (k - s_{3}) \cos \theta \over \sin \theta } \; ] \;
\vec{E} =   m \; \Phi_{0} \; ,
\nonumber
\\
- \partial_{t} \; \vec{E} \; - \;
(\partial_{r} +  {1 \over r}) \; \tau_{3} \; \vec{H} \; + \; \sigma \;
\vec{e}^{\;t}_{3} \partial_{r} C \; +
 \sigma \; {1 \over r} \;
[ \vec{e}^{\;t}_{1} \; \partial_{\theta}   \; +  \;
 \; \vec{e}^{\;t}_{2}(-i)  { -m  -  k \cos \theta  \over
\sin \theta } ] \; C  -
\nonumber
\\
- \;
{1 \over r} \; [ \tau_{1} \; \partial_{\theta} +
\tau_{2} (-i) \; { - m  -  (k - s_{3} ) \cos \theta \over \sin \theta } ] \;
\vec{H} =  m \; \vec{\Phi}  \; ,
\nonumber
\\
\partial_{t} \vec{\Phi} \; - \; \vec{e}^{\;t}_{3} \;
\partial_{r} \; \Phi_{0} \; - \;
\; {1 \over r} \;
[\;  \vec{e}^{\;t}_{1} \; \partial_{\theta}   \; +  \;
 \; \vec{e}^{\;t}_{2} (-i) \; { - m  - k \cos \theta  \over
\sin \theta } \; ] \; \Phi _{0} =   m  \; \vec{E} \; ,
\nonumber
\\
(\partial_{r} + {1 \over r}) \; \tau_{3} \; \vec{\Phi} \; + \;
{1 \over r} \; [ \; \tau_{1} \; \partial_{\theta} +
\tau_{2} (-i)  \; { - m  -  (k - s_{3}) \cos \theta \over \sin \theta } \; ] \;
\vec{\Phi} =  m \; \vec{H} \;
\label{4.12}
\end{eqnarray}

\noindent where
$$
s_{3} = \left | \begin{array}{ccc}
+1  &  0  &  0  \\
0   &  0  &  0  \\
0   &  0  & -1  \end{array} \right | \;
$$

\noindent with the use of several intermediate formulas
\begin{eqnarray}
\partial_{\theta} \vec{C} =
e^{-i\epsilon t }
\left |  \begin{array}{c}
C_{1} \partial_{\theta} D_{k-1}  \\
C_{2} \partial_{\theta} D_{k}    \\
C_{3} \partial_{\theta} D_{k+1}  \end{array}  \right |
=
e^{-i\epsilon t }
\left | \begin{array}{c}
C_{1} ( a \;  D_{k-2} - c \; D_{k}   )  \\
C_{2} ( c \;  D_{k-1} - d \; D_{k+1} )  \\
C_{3} ( d \;  D_{k}   - b \; D_{k+2} )      \end{array}  \right | \; ,
\nonumber
\\
{ - m -  (k - s_{3}) \cos \theta \over \sin \theta }  \; \vec{C } =
 e^{-i \epsilon t}
\left | \begin{array}{c}
C_{1} ( - a \;  D_{k-2} - c \; D_{k}   )  \\
C_{2} ( - c \;  D_{k-1} - d \; D_{k+1} )  \\
C_{3} ( - d \; D_{k}   - b \; D_{k+2} )  \end{array} \right | \; ,
\nonumber
\end{eqnarray}
\begin{eqnarray}
\vec{e}_{1}  \; \partial_{\theta} \; \vec{C} =
 e^{-i\epsilon t } \; {1 \over \sqrt{2}} \;  [ \;-
C_{1} ( a \;  D_{k-2} - c \; D_{k}   )\;
+\; C_{3} \; ( d \;  D_{k}   - b \; D_{k+2} ) \; ] \; ,
\nonumber
\\
- i \vec{e}_{2}\;
{ - m -  (k - s_{3}) \cos \theta \over \sin \theta }  \; \vec{C} =
\nonumber
\\
= {e^{-i \epsilon t}  \over \sqrt{2}}
[ \; - C_{1}\; ( - a \;  D_{k-2} - c \; D_{k}   ) \; -\;
C_{3} \;( - d \; D_{k}   - b \; D_{k+2} ) \; ] \; ,
\nonumber
\\
( \; \vec{e}_{1}  \partial_{\theta}
- i \; \vec{e}_{2}
{ - m -  (k - s_{3}) \cos \theta \over \sin \theta }  \; ) \; \vec{C}=
e^{-i\epsilon t } \; \sqrt{2}  \;(  \; c \; C_{1} \;+ \; d \; C_{3}\; )\; D_{k} \; .
\nonumber
\\
\tau_{1} \; \partial_{\theta} \; \vec{H} \; +\;
\tau_{2}\; (-i) \; { -m  -  (k - s_{3} ) \cos \theta \over \sin \theta }  \;\; \vec{H} =
 e^{-i \epsilon t}  \;(  i \sqrt{2} ) \;
\left |  \begin{array}{c}
                       - c\; H_{2} \; D_{k-1}  \\
                       (c \; H_{1} - d \; H_{3})\; D_{k}  \\
                        + d\;  H_{2} \; D_{k+1}            \end{array} \right | \; .
\nonumber
\\
(\;  \vec{e}^{\;\;t}_{1} \; \partial_{\theta}   \; +  \;
 \; \vec{e}^{\;\;t}_{2} (-i) \; { - m  - k  \cos \theta \over
 \sin \theta } \; ) \;C   =   e^{-i \epsilon t}  \; \sqrt{2} \;
\left |  \begin{array}{c}
                           - c\; C(r) \; D_{k-1}  \\
                                     0            \\
                           - d \; C(r) \; D_{k+1}      \end{array} \right |.
\nonumber
\end{eqnarray}

\noindent
from (\ref{4.12}) you arrive  at
\begin{eqnarray}
- i \epsilon \; C_{0} - ( {d \over d r} + {2 \over r}) \; C_{2}  - {\sqrt{2} \over r}
\; (c \;  C_{1} + d \; C_{3} ) = m \; C \; ,
\label{4.13a}
\end{eqnarray}
\begin{eqnarray}
\qquad  \;\;\;  - ( {d \over d r} + {2 \over r}) \;  E_{2} -
{\sqrt{2} \over r} \; ( c \;  E_{1}  + d \; E_{3}) = m\; C_{0} \; ,
\nonumber
\\
+i \epsilon \; E_{1}  + i ( {d \over d r } + {1 \over r}) \; H_{1}  \qquad
 +{i\sqrt{2} \; c \over r} \;  H_{2}= m \; C_{1} \; ,
\nonumber
\\
+i \epsilon E_{2}    \qquad  -  {i\sqrt{2} \over r} \;
(c\; H_{1} - d \;  H_{3}) = m \; C_{2} \; ,
\nonumber
\\
+i \epsilon  \; E_{3}  - i ( {d \over d r } + {1 \over r}) \; H_{3}    \qquad  -
{i\sqrt{2}\; d  \over r} \;  H_{2}= m \;  C_{3} \; ,
\label{4.13b}
\end{eqnarray}
\begin{eqnarray}
-i \epsilon \; \sigma \; C -  ( {d \over d r} + {2 \over r} )\;  E_{2}  -
 { \sqrt{2} \over r}
(c \; E_{1} + d \; E_{3})  = m  \; \Phi_{0} \; ,
\nonumber
\\
+i \epsilon  \; E_{1}  + i ( {d \over d r } + {1 \over r}) \;  H_{1} - \sigma \; {\sqrt{2}\; c \over r} \;
  C\;\;\; + {i\sqrt{2} c \over r} \;  H_{2} = m  \; \Phi_{1} \; ,
\nonumber
\\
+i \epsilon \; E_{2}    + \sigma \; {d \over d r} C \;\;  -  {i\sqrt{2} \over r}
(c \; H_{1} - d \; H_{3}) = m \; \Phi_{2} \; ,
\nonumber
\\
+i \epsilon \; E_{3}  - i ( {d \over d r } + {1 \over r}) \; H_{3}    \;
- \sigma \; { \sqrt{2}\; d \over r} \;  C  - {i\sqrt{2}\; d \over r} \;  H_{2}= m \; \Phi_{3} \; ,
\label{4.13c}
\end{eqnarray}
\begin{eqnarray}
-i \epsilon \; \Phi_{1} \;   + {\sqrt{2}\; c \over r} \;  \Phi_{0}  = m \; E_{1} \; ,
\;\;
- i \epsilon \;  \Phi_{2} - {d \over d r} \; \Phi_{2} \qquad = m \;  E_{2} \; ,
\nonumber
\\
-i \epsilon \; \Phi_{3} + {\sqrt{2}\;d \over r }  \; \Phi_{0}  = m  \; E_{3} \; ,
\;\;
-i ({d \over d r } + {1 \over r}) \;  \Phi_{1} - {i \sqrt{2}\; c   \over r } \; \Phi_{2} = m \; H_{1} \; ,
\nonumber
\\
{ i \sqrt{2} \over r} \;  (c \; \Phi_{1} - d \;  \Phi_{3}) = m \; H_{2} \; ,
\;\;
+i ({d \over d r } + {1 \over r} ) \; \Phi_{3} + {i \sqrt{2}\; d   \over r } \;  \Phi_{2} = m \; H_{3} \; .
\label{4.113d}
\end{eqnarray}

\section{States with $j=j_{min} $ in monopole field}

\hspace{5mm}
As known, in  the monopole problem there  arises a very peculiar situation for states with
minimal value of $j$. Now  we are going to consider those in our case.
In accordance  with quantization conditions by Dirac-Schwinger
the parameter  $k = eg / \hbar c $ involved will take on the  values
\begin{eqnarray}
k = \pm1 , \pm 3/2, \pm 2, ...
\qquad\mbox{and} \qquad j = \mid k \mid - 1 , \mid k \mid , \mid k \mid +1 , ...
\label{5.1}
\end{eqnarray}

One should consider with more details several particular cases.
Firstly, let $k$ be
$$
k = +1,  \; \; j_{min} = 0 \; .
$$

\noindent
Correspondingly,  a wave function should be taken in the form
\begin{eqnarray}
\underline{\Psi^{j=0}_{k=+1}(x)}\; :  \;\;\; C(r)  = 0  \; , \;\;\;\;
C_{0}(r) =  0 \; , \;\;
\Phi_{0} (r) = 0  \; ,
\nonumber
\\
\vec{C} (x) =  e^{-i\epsilon t} \;
\left  | \begin{array}{c}
C_{1}(r)   \\  0   \\  0
\end{array} \right | \; , \;
\vec{\Phi}(x)   = e^{-i\epsilon t}\;
\left |  \begin{array}{c}
\Phi_{1} (r)  \\  0   \\  0 \end{array} \right | \; ,
\nonumber
\\
\vec{E} (x)  = e^{-i\epsilon t} \;
\left |  \begin{array}{c}
E_{1}(r)   \\  0  \\   0
\end{array} \right | \; , \;
\vec{H} (x) = e^{-i\epsilon t} \;
\left |  \begin{array}{c}
H_{1}(r)  \\  0  \\  0 \end{array} \right | \; .
\label{4.18b}
\end{eqnarray}

\noindent On those $\Psi^{j=0}(x)$ the angular operator vanishes identically
$
\Sigma^{k}_{\theta, \phi} \Psi^{j=0}(x) = 0 \; ,
$
and corresponding radial equations are as follows
\begin{eqnarray}
+i \; \epsilon \; E_{1} + i\; ({d \over d r} + {1 \over r}) \; H_{1} = m \; C_{1} \; ,
\nonumber
\\
+i\; \epsilon \; E_{1} + i \;({d \over d r} + {1 \over r})\; H_{1} = m \; \Phi_{1} \; ,
\nonumber
\\
\; - i \; \epsilon \;  \Phi_{1} = m \;  E_{1} \; , \; - i\; ({d \over d r}  + {1 \over r})\;
 \Phi_{1} = m \; H_{1} \; .
\nonumber
\end{eqnarray}

\noindent
Excluding $H_{1},E_{1}$ for $C_{1}=\Phi_{1}$ one gets
\begin{eqnarray}
( { d^{2} \over dr^{2} } + \epsilon^{2} - m^{2} )\;  { \Phi_{1} \over r} = 0 \;
\qquad
\Longrightarrow \qquad
{\Phi _{1} \over r } = \exp \;[\; \pm \sqrt{m^{2} - \epsilon^{2}} \; ] \; .
\label{4.19b}
\end{eqnarray}

\noindent
One of these solutions  might be associated with
a "bound state", this is what makes them very interesting from physical standpoint.

Another variant
\begin{eqnarray}
k = - 1 ,\;\;  j_{min} = 0
\nonumber
\end{eqnarray}

\noindent will  look the same. Now an initial substitution is
\begin{eqnarray}
\underline{\Psi^{j=0}_{k=+1}(x) \;} :  \;\;\; C(r)  = 0  \; , \;\;\;\;
C_{0}(r) =  0 \; , \;\;
\Phi_{0} (r) = 0  \; , \;\;
\nonumber
\\
\vec{C} (x) =  e^{-i\epsilon t} \;
\left |  \begin{array}{c}
  0 \\  0   \\ C_{3}(r)
\end{array} \right | \; , \;
\vec{\Phi}(x)   = e^{-i\epsilon t}\;
\left |  \begin{array}{c}
0   \\  0   \\  \Phi_{3} (r)
 \end{array} \right | \; , \;
\nonumber
\\
\vec{E} (x)  = e^{-i\epsilon t} \;
\left |  \begin{array}{c}
0  \\  0  \\  E_{3}(r)
\end{array} \right |  \; , \;
\vec{H} (x) = e^{-i\epsilon t} \;
\left |  \begin{array}{c}
0   \\  0  \\  H_{3}(r) \end{array} \right | \; .
\label{4.20b}
\end{eqnarray}

\noindent Again on those functions $\Psi^{j=0}(x)$,
$
\Sigma^{k}_{\theta, \phi} \; \Psi^{j=0}(x) = 0 \;,
$
and  radial system will be
\begin{eqnarray}
+i\; \epsilon \; E_{3} - \;i\; ({d \over d r} + {1 \over r})\; H_{3} = m \; C_{3} \; ,
\;
\nonumber
\\
+i\; \epsilon \; E_{3} + i\; ({d \over d r} + {1 \over r}) \; H_{3} = m \; \Phi_{3} \; ,
\nonumber
\\
\; - i \; \epsilon \; \Phi_{3} = m\; E_{3} \; , \;  i \;({d \over d r}  + {1 \over r})\; \Phi_{3} = m \; H_{3} \; .\,
\nonumber
\end{eqnarray}

\noindent
which after excluding   $H_{3},E_{}$ gives
\begin{eqnarray}
( { d^{2} \over dr^{2} } + \epsilon^{2} - m^{2} )\;  { \Phi_{3} \over r} = 0
\qquad \Longrightarrow \qquad
{\Phi _{3} \over r } = \exp [\; \pm \sqrt{m^{2} - \epsilon^{2}} \; ] \; .
\label{4.21b}
\end{eqnarray}

Now  we consider states with minimal
$j_{min} = \mid k \mid -1$  when $k= \pm3/2, \pm 2, \pm5/2, ...$
Here, though the $\theta,\phi$-variables enters  $\Psi^{j_{min}}(x)$,
however the angular operator acts on these functions as zero-like.
Actually, as
\begin{eqnarray}
k = +3/2, + 2, + 5/2, ... \;, \;\;\;  j_{min} = \mid k \mid -1 \; ,
\nonumber
\\[2mm]
\Psi^{j_{min}}(x) \; : \;\;\; C(r)  = 0  \; , \;\;\;\;
C_{0}(r) =  0 \; , \;\;
\Phi_{0} (r) = 0  \; ,
\nonumber
\\
\vec{C} (x) =  e^{-i\epsilon t} \;
\left | \begin{array}{c}
C_{1}(r) D_{k-1}  \\  0   \\  0
\end{array} \right | \; , \;
\vec{\Phi}(x)   = e^{-i\epsilon t}\;
\left |  \begin{array}{c}
\Phi_{1} (r) D_{k-1} \\  0   \\  0 \end{array} \right | \; ,
\nonumber
\\
\vec{E} (x)  = e^{-i\epsilon t} \;
\left |  \begin{array}{c}
E_{1}(r) D_{k-1}  \\  0  \\   0
\end{array} \right | \; , \;
\vec{H} (x) = e^{-i\epsilon t} \;
\left |  \begin{array}{c}
H_{1}(r) D_{k-1} \\  0  \\  0 \end{array} \right | \;  .
\label{4.22}
\end{eqnarray}

\noindent Here you need the recurrence formulas [2]
$$
\partial_{\theta} D_{k-1} = \sqrt{{k -1 \over 2}} D_{k-2} \; ,\qquad
{ - m - (k-1) \cos \theta  \over \sin \theta } D_{k-1} =
-\sqrt{{k -1 \over 2} } D_{k-2} \; .
$$

\noindent with the help of  which you easy can prove
$
\Sigma_{\theta,\phi} \; \Psi^{j_{min}}(x) = 0 \;
$
and  then again  we  arrive at  eqs.  (\ref{4.19b}).

Another variant
\begin{eqnarray}
k = -3/2, - 2, - 5/2, ... \;, \;\;\;  j_{min} = \mid k \mid -1 \; ;
\nonumber
\end{eqnarray}

\noindent looks the same:
\begin{eqnarray}
\Psi^{j_{min}}(x) \; : \;\;\; C(r)  = 0  \; , \;\;\;\;
C_{0}(r) =  0 \; , \;\;
\Phi_{0} (r) = 0  \; ,
\nonumber
\\[2mm]
\vec{C} (x) =  e^{-i\epsilon t} \;
\left |  \begin{array}{c}
0   \\  0   \\  C_{3}(r) D_{k+1}
\end{array} \right | \; , \;
\vec{\Phi}(x)   = e^{-i\epsilon t}\;
\left |  \begin{array}{c}
0  \\  0   \\  \Phi_{3} (r) D_{k+1} \end{array} \right | \; ,
\nonumber
\\
\vec{E} (x)  = e^{-i\epsilon t} \;
\left |  \begin{array}{c}
0   \\  0  \\   E_{+}(r) D_{k+1}
\end{array} \right | \; , \;
\vec{H} (x) = e^{-i\epsilon t} \;
\left |  \begin{array}{c}
0  \\  0  \\  H_{3}(r) D_{k+1} \end{array} \right | \; ,
\label{4.23b}
\end{eqnarray}
\begin{eqnarray}
\partial_{\theta} D_{k+1} = - \sqrt{-{(k + 1) \over 2}} D_{k+2} \; ,\;
\nonumber
\\
{ - m - (k+1) \cos \theta  \over \sin \theta } D_{k+1} =
- \sqrt{-{(k + 1) \over 2} } D_{k+2} \; ,
\nonumber
\end{eqnarray}

\noindent with the help of  which you easy can prove
$
\Sigma_{\theta,\phi} \; \Psi^{j_{min}}(x) = 0 \;
$
and  then again  we  arrive at  eqs.  (\ref{4.21b}).

\section{On the  Coulomb problem of an ordinary $S=1$ particle}

\hspace{5mm}
Quantum mechanical study of a vector particle in Coulomb field has a long history.
We are going to review this old problem else one time in the frame of the approach developed
above. So, let us consider a 10-component Duffin-Kemmer equation for a S=1 particle in Coulomn
field all  details on the working technic used here and based on the tetrad
formalism).

In the spherical-cyclic basis the most general form of 10-component wave function with quantum numbers
$\epsilon , j , m$ is
\begin{eqnarray}
\Psi (x) =
\{  \;  \Phi_{0} (x) ,
\; \vec{\Phi}(x) , \; \vec{E}(x) , \vec{H}(x) \; \} \; ,
\nonumber
\\
\Phi_{0}(x) =    e^{-i\epsilon t}\; \Phi_{0} (x) \; D_{0} \; ,
\;
\vec{\Phi}(x)   = e^{-i\epsilon t}\;
\left |  \begin{array}{c}
\Phi_{1} (r)\; D_{-1}  \\
\Phi_{2}(r) \; D_{0}   \\
\Phi_{3}(r) \; D_{+1}  \end{array} \right | \; ,
\nonumber
\\
\vec{E}   = e^{-i\epsilon t} \;
\left |  \begin{array}{c}
E_{1}(r) \;  D_{-1} \\
E_{2}(r) \;  D_{0}   \\
E_{3}(r) \;  D_{+1}     \end{array} \right | \; , \;\;
\vec{H}  = e^{-i\epsilon t} \;
\left |  \begin{array}{c}
H_{1}(r)  \; D_{-1}  \\
H_{2}(r)  \; D_{0}   \\
H_{3}(r)  \; D_{+1}    \end{array} \right |\; .
\label{6.1}
\end{eqnarray}

\noindent    where
$
D_{s}=
D_{-m,s}^{j}(\phi,\;\theta,\;0) \; , \;\; s = 0,\;+1,\;-1 \; .
$
Omitting details of separation of variables let us  immediately write down a radial system
( that can be derived directly from 15-component above by setting $\sigma = 0$ and so on)
\begin{eqnarray}
 - \;\;({d \over dr}  + \frac {2}{r}) \; E_{2} -
\frac{\nu}{r}\;  (E_{1} + E_{3}) = m \Phi_{0}  \; ,
\nonumber
\\
+ i (\epsilon + {\alpha \over r} ) \; E_{1} \;\; + i \;({d \over dr} + \frac {1}{r}) \; H_{1} +
i \frac{\nu}{r} \; H_{2} =  m \Phi_{1} \; ,
\nonumber
\\
+ i ( \epsilon + {\alpha \over r}) \; E_{2} \;\;  -
i\frac{\nu}{r}\; (H_{1} - H_{3})
=  m \Phi_{2}  \; ,
\nonumber
\\
+i(\epsilon + {\alpha \over r}) \; E_{3} \;\; - i \;({d\over dr} + \frac {1}{r})\; H_{3} -
i \frac{\nu}{r} \; H_{2} =  m \; \Phi_{3}  \;   ,
\label{(6.2a}
\end{eqnarray}
\begin{eqnarray}
-i (\epsilon + {\alpha \over r}) \; \Phi_{1}\;\;  +
\frac{\nu}{r} \; \Phi_{0} - m E_{1} = 0 \;     , \;
\nonumber
\\
-i (\epsilon +{\alpha \over r})  \; \Phi_{2} \;\; - \;{d \over  dr} \; \Phi_{0}
- m E_{2} = 0 \;    ,
\nonumber
\\
-i ( \epsilon  + {\alpha \over r} ) \; \Phi_{3} \;\;  +
\frac{\nu}{r} \; \Phi_{0}  \; - \; m E_{3} = 0 \;   , \;
\nonumber
\\
 - i\; ({d \over dr} + \frac {1}{r}) \; \Phi_{1} -
i\frac{\nu}{r} \; \Phi_{2} \;  - \; m H_{1} = 0 \;  ,
\nonumber
\\
+i\frac{\nu}{r}\; (\Phi_{1} - \Phi_{3}) - m H_{2} = 0 \; ,
\nonumber
\\
  +i\;({d \over dr} + \frac {1}{r})\; \Phi_{3} +
i\frac{\nu}{r} \Phi_{2} - m H_{3} = 0 \; .
\label{6.2b}
\end{eqnarray}

\noindent
Concurrently with  $\vec{J}^{\;2}, J_{3}$ one may diagonalize  the operator
of the space inversion  $\hat{\Pi}$.
That  in chosen representation has the form
\begin{eqnarray}
\hat{\Pi}   =  [\;( 1 \oplus \Pi_{3}) \oplus ( \Pi_{3} \oplus -\Pi_{3})\; ]\;
\hat{P} \; , \;\;\;
\Pi_{3}  =    \left |  \begin{array}{rrr}
0  &  0  &  -1  \\
0  &  -1 &   0  \\
-1  &  0  &  0   \end{array} \right | \; .
\nonumber
\end{eqnarray}

\noindent The eigenvalue equation  $\hat{\Pi} \Psi = P \; \Psi $
gives solutions of two types:
\begin{eqnarray}
\underline{P=(-1)^{j+1}} , \;\;\qquad
\Phi_{0} = 0 \; , \;\;  \Phi_{3} = - \Phi_{1} \; , \;\;
\Phi_{2} = 0 \; ,
\nonumber
\\
E_{3} = - E_{1} \; , \; E_{2} = 0 , \; H_{3}  = H_{1} \; ;
\nonumber
\\[2mm]
\underline{P=(-1)^{j}} \; , \qquad
E_{3} = + E_{1} \; , \; H_{3} = - H_{1} \; , \; \;
H_{2} = 0 \; .
\label{P}
\end{eqnarray}

\noindent
As for the case $ P=(-1)^{j+1} $ you have the system
\begin{eqnarray}
+i ( \epsilon  + {\alpha \over r } ) \; E_{1} +  i ({d \over dr} +{ 1 \over r}) H_{1} +
 i {\nu \over r} H_{2}
= m \Phi _{1} \; ,
\nonumber
\\
-i  ( \epsilon  + {\alpha \over r}) \; \Phi_{1}  = m E_{1} \; , \;\;
-i ({d \over dr} + {1 \over r}) \Phi_{1}   = m H_{1} \; , \;\;
2i { \nu \over r }  \Phi_{1} = m H_{2}  \; ,
\nonumber
\end{eqnarray}

\noindent
which after excluding  $E_{1}, H_{1}, H_{2}$
leads to
\begin{eqnarray}
\left [ \;{ d^{2} \over dr^{2} } + {2 \over r} {d \over d r}  + (\epsilon + {\alpha \over r})^{2} -
{j(j+1) \over r^{2}} \; \right ] \; \Phi_{1} = 0 \; .
\label{6.5a}
\end{eqnarray}

\noindent
This exactly the same equation that arises in the theory of a scalar particle in Coulomb field.
Its energy spectrum and wave functions are well known.

States with  $P=(-1)^{j}$  are characterized by the system
\begin{eqnarray}
({d \over dr} +{2\over r}) E_{2} -
2 {\nu \over r} E_{1} = m \Phi_{0}\; ,
\nonumber
\\
+ i ( \epsilon + {\alpha \over r}) \;  E_{1} + i ( {d \over dr} + {1 \over r}) H_{1}  =
m \Phi_{1} \; ,
\nonumber
\\
+ i ( \epsilon + {\alpha \over r})  E_{2} -2i {\nu \over r}  H_{1} =
m \Phi_{2} \; ,
\nonumber
\\
\;\;
-i ( \epsilon + {\alpha \over r}) \; \Phi_{1} + {\nu \over r} \Phi_{0} = m E_{1} \; ,
\nonumber
\\
-i ( \epsilon + {\alpha \over r}) \Phi_{2}  - {d \over dr} \Phi_{0} = m E_{2} \; ,
\nonumber
\\
-i ({d \over dr} +{1\over r}) \Phi_{1} -i {\nu \over r} \Phi_{2} = m H_{1} \;
\label{6.5b}
\end{eqnarray}

\noindent
in solving which we have not been able to succeed.

The above radial equations (\ref{6.1}) as well as substitutions (\ref{6.1}) are
correct only for $j=1, 2,...$
However you should consider the  $j=0$ case separately and  having started  with
a special wave function:
\begin{eqnarray}
C_{0}(x) =e^{-i\epsilon t} C_{0}\; (r) , \qquad
\Phi_{0}(x) =  e^{-i\epsilon t}  \Phi_{0} (r) \; ,
\nonumber
\\
\vec{C} (x) = e^{-i\epsilon t} \left |  \begin{array}{c}
0  \\
 C_{2}(r)     \\
0  \end{array} \right | \; , \;
\vec{\Phi}(x)   = e^{-i\epsilon t}
\left |  \begin{array}{c}
0   \\
\Phi_{2}(r)    \\
0  \end{array} \right | \; ,
\nonumber
\\
\vec{E} (x)  = e^{-i\epsilon t}
\left |  \begin{array}{c}
0  \\
E_{2}(r)   \\
0  \end{array} \right | \; , \;\;
\vec{H} (x) =  e^{-i\epsilon t} \left |  \begin{array}{c}
0  \\
H_{2}(r)  \\
0    \end{array} \right |\; .
\label{5.6}
\end{eqnarray}

\noindent
This $\Psi$ is eigenfunction of $\hat{\Pi}$ with $\Pi=+1$.
Corresponding radial system is
($\Phi_{0} = \varphi_{0}, -i  \Phi_{1} = \varphi_{1}, -i \Phi_{2} = \varphi_{2}$)
\begin{eqnarray}
H_{2} = 0\; , \qquad - ({d \over d r} +{2 \over r}) E_{2} = m \varphi_{0}\; ,
\nonumber
\\
(\epsilon  + {\alpha \over r})  E_{2}   = m \varphi_{2} \; , \qquad
(\epsilon + {\alpha \over r}) \varphi_{2}  - {d \over d r} \varphi_{0} = m E_{2} \; ,
\nonumber
\end{eqnarray}

\noindent
from where it follows (that corresponds to  (\ref{3.9}) when $\sigma = 0$)
\begin{eqnarray}
\left [ \; { d^{2} \over dr^{2}} +
{2 \over r} {d \over d r} - {2 \over r^{2}}+ ( \epsilon + {\alpha \over r})^{2} - m^{2} \right ]\;
 E_{2} = 0 \; ,
\nonumber
\end{eqnarray}

\noindent  or introducing $E_{2}(r) = r^{-1} f(r)$
\begin{eqnarray}
{d^{2} \over dr^{2}} \; f +   ( \; \epsilon^{2} - m^{2} + {2 \alpha \epsilon \over r} -
 {2 - \alpha^{2} \over r^{2}} \;  )\;   f = 0 \; .
\label{6.9}
\end{eqnarray}

\noindent  The latter will read in usual dimension units as
\begin{eqnarray}
{d^{2} \over dr^{2}} f + \left [\;  {E^{2} \over c^{2}  \hbar^{2}}
 - {M^{2} c^{2} \over \hbar^{2}}  + { 2 Z (e^{2} /c \hbar) (E /c \hbar)  \over r} -
  {2 - Z^{2} (e^{2} / c \hbar)^{2}  \over r^{2}} \;  \right ] \;  f = 0 \; .
\nonumber
\end{eqnarray}

\noindent
Now you can pass to a dimensionless coordinate $ x = r E /  c \hbar $, then the equation
becomes
\begin{eqnarray}
{d^{2} \over dx^{2}} f +
\left [ \; 1 - { M^{2} c^{4} \over E^{2}} +
 2 Z  (e^{2} /c \hbar ) \; {1 \over x} -
{ 2 -  Z^{2} ( e^{2} / c \hbar  ) ^{2} \over x^{2} }  \; \right ] \; f (x) = 0
\; .
\nonumber
\end{eqnarray}

\noindent
With the notation
\begin{eqnarray}
{ M^{2} c^{4} \over E^{2}} = \Lambda^{2} \; , \qquad
Z \; {e^{2} \over c \; \hbar}  = Z \; {1 \over 137} = \gamma \;  < \;1
\nonumber
\end{eqnarray}

\noindent the equation will read as
\begin{eqnarray}
{d^{2}  \over dx^{2} } \; f\;  + \;
( \; 1 \; - \; \Lambda ^{2} \;  + \;
{ 2 \gamma  \over x} \; - \; { 2 -  \gamma^{2} \over x^{2} }  \; ) \; f  = 0 \; .
\label{5.13}
\end{eqnarray}

\noindent
To obtain bound states you should take   $f(x)$ in the form
\begin{eqnarray}
f(x) = x^{a}\;  e^{-bx} \;F (x) \;
\nonumber
\end{eqnarray}

\noindent where $a$ and $b$ are expected to be positive whereas  $F(x)$
be a polynomial in $x$. From  (\ref{5.13}) we derive
\begin{eqnarray}
 x \;  F'' \; +  \; ( 2a - 2bx) \;  F' \;  + \qquad\qquad
\nonumber
\\
 + \;  [ \;
{a(a-1) + \gamma^{2} - 2 \over x} + (b^{2} +1 - \Lambda^{2} ) x + (2\gamma - 2ab ) \;  ]\;  F = 0 \;.
\nonumber
\end{eqnarray}

\noindent This equation  with demands
\begin{eqnarray}
a(a-1) + \gamma^{2} - 2 = 0 \; , \qquad
b^{2} +1 - \Lambda^{2} = 0\;
\nonumber
\end{eqnarray}

\noindent leads to
$$
 x \;  F'' \; +  \; 2 ( a - b x) \;  F' \; + \;
2(\gamma - a b ) \;  F = 0 \; .
\eqno(5.15c)
$$

\noindent
For $a$ and $b$ we have
\begin{eqnarray}
a = {1 \pm \sqrt{9 - 4 \gamma^{2}} \over 2} \; ,
\qquad
b = \pm \sqrt{\Lambda^{2} - 1} = \pm { \sqrt{M^{2} c^{4} - E^{2}} \over E } \; .
\label{6.16}
\end{eqnarray}

\noindent The choice of "upper" \hspace{2mm} signs is what you need to  $a$ and $b$ be positive.

Further, taking $F(x)$ as a series
\begin{eqnarray}
F (x) = \sum_{k=0}^{\infty}\; C_{k} \; x^{k} \; ,
\;\;
F' = \sum_{k=1}^{\infty} \; k C_{k}\;  x^{k-1} \; , \;\;
F'' = \sum_{k=2}^{\infty} \; k(k-1) C_{k} \; x^{k-2} \; ,
\nonumber
\end{eqnarray}

\noindent you get
$$
\sum_{k=2}^{\infty} \; k (k-1) C_{k}\; x^{k-1} \; + \;
2a \; \sum_{k=1}^{\infty} \; k  C_{k} \; x^{k-1} \; - \;
2b  \; \sum_{k=1}^{\infty} \; k  C_{k} \; x^{k} \; + \;
2(\gamma - a b) \;  \sum_{k=0}^{\infty} \; c_{k} \; x^{k} = 0 \;
$$

\noindent or
\begin{eqnarray}
\sum_{n=1}^{\infty} \; n (n+1) C_{n+1} \; x^{n} \; + \;
2a \; \sum_{n=0}^{\infty} \; (n+1)  C_{n+1} \; x^{n} \; - \;
\nonumber
\\
- \;2b  \; \sum_{n=1}^{\infty} \; n  C_{n} \; x^{n} \; + \;
2(\gamma - a b) \;  \sum_{n=0}^{\infty} \; C_{n} \; x^{n} = 0 \;
\nonumber
\end{eqnarray}

\noindent or
\begin{eqnarray}
[ \; 2aC_{1} + 2 (\gamma - a b) \; ] x^{0} \; + \;
[ \; 2 C_{2} + 2a \; 2C_{2} - 2 b \; C_{1} + 2(\gamma -a b) C_{1} \; ] \; x +
\nonumber
\\
+ \;
\sum_{n=2}^{\infty} \; [ \;
n(n+1) C_{n+1} + 2a (n+1) C_{n+1} - 2bnC_{n} + 2 (\gamma -a b ) C_{n} \; ] \; x^{n} = 0 \; .
\nonumber
\end{eqnarray}

\noindent
Demanding  coefficients at all $x^{k}$ be equal to zero, we produce recursive relations
\begin{eqnarray}
C_{1} = - (\gamma -a b) \; C_{0} = 0 \; ,
\nonumber
\\
 C_{2} \;2 \;  (1  + 2 a )  = 2 \; [ b - (\gamma -a b)] \; C_{1} = 0
, \;  n = 2,3,4, ... ,
\nonumber
\\
C_{n+1} \; (n+1)\;  (  n + 2a )  =  2 \; [ n\; b\;  -\;   (\gamma -a b )\; ]\; C_{n}   = 0 \; .
\label{5.17c}
\end{eqnarray}

\noindent
To terminate an infinite series at a certain place you need impose
\begin{eqnarray}
C_{N+1} = 0 \; \Longrightarrow \; \; [\; N \; b\;  - \;   (\gamma - a b )\; ] \; =0 \;
\nonumber
\end{eqnarray}

\noindent
that provides us with  a quantization condition to produce  an energy spectrum.
It looks as
\begin{eqnarray}
{ \gamma - a b \over b } = N \;
\nonumber
\end{eqnarray}

\noindent which having remembered   $a$ and $b$  (5.16) will take the form
(the notation $2\Gamma = (1 + \sqrt{9 -4\gamma^{2}}$ is used) ,
\begin{eqnarray}
{ 2 \gamma \epsilon - \Gamma  \sqrt{m^{2}c^{4} - \epsilon^{2} }\over
2 \sqrt{m^{2}c^{4} - \epsilon^{2} }} = N
\nonumber
\end{eqnarray}

\noindent from which you arrive  at the energy formula
\begin{eqnarray}
\epsilon = mc^{2} \left [ 1 + {\gamma^{2}  \over (\Gamma + N )^{2}  }  \; \right ] ^{-1/2} \; .
\label{6.20}
\end{eqnarray}

Some additional remarks might be given. The material of this Section  might be of some interest in the
light of the well-known  old results obtained by I.E. Tamm [...]  on behavior of
the vector particle wave  functions in presence of external Coulomb field.

The most principal part of this work consists in the following. All possible solutions can be
divided into two classes. One is reduced to the well known functions  of the scalar particle in Coulomb
field with usual and expected  behavior  at the origin $r=0$ and at infinity $r = \infty $.
However   another part of wave  functions  provides us  with  big surprise -- they are singular at the
origin, corresponding differential equations have not been resolved up to now; it is generally
accepted that they should be  associated with a situation of
falling down to centre $r=0$.

In essence, results of this Section prove that there exists definite correlation between
values of  $j$, of  $P$, and property of wave function to be singular or non-singular;
it can be seen  in the following table:

\vspace{5mm}

\begin{tabular}{|c|c|c|c|c|}
\hline
  Value          & $P=(-1)^{j}$&$P=(-1)^{j}$   &$P=(-1)^{j+1}$ & $P=(-1)^{j+1}$ \\
  of quantum number & singular  & non-singular & cingular   & non-singular  \\
                   & solution     & solution        & solution      & solution   \\
\hline
  $j=0$            & ---         & $P=+1$        & ---         & --- \\
  $j=1$            & $P=-1$      & ---           & ---         & $P=+1$ \\
  $j=2$            & $P=+1$      & ---           & ---         & $P=-1$ \\
  $j=3$            & $P=-1$      & ---           & ---         & $P=+1$ \\
  $j=4$            & $P=+1$      & ---           & ---         & $P=-1$ \\
  ...              &  ...        & ...           & ...         & ... \\ \hline
\end{tabular}


\end{document}